\documentclass[10pt]{article}

\usepackage[utf8]{inputenc}
\usepackage[english]{babel}
\usepackage{ragged2e}
\usepackage{graphicx}
\usepackage{float}
\usepackage{amsthm}
\usepackage{mathtools}
\usepackage{amsfonts}
\usepackage{dsfont} 
\usepackage{amssymb}
\usepackage{bbm}
\usepackage{bm}
\usepackage{tikz}
\usepackage{algorithm}
\usepackage{relsize}
\usepackage{algorithm,algpseudocode}
\usepackage{breqn}
\usepackage[hyphens]{url}
\usepackage{etoolbox}
\usepackage{lmodern}
\usepackage{natbib}
\usepackage{tikz}
\usepackage[colorlinks=true,linkcolor=blue, citecolor=blue]{hyperref}
\usepackage{mathtools}

\mathtoolsset{showonlyrefs}
\usepackage{subcaption}
\usepackage{subfloat}
\usepackage{letltxmacro}%
\usepackage{scalerel}
\usepackage[toc,page]{appendix}
\usepackage{tabularx,ragged2e,booktabs,caption}
\usepackage[font=small,labelfont=bf]{caption}
\usepackage{blindtext}
\usepackage[%
    font={small,sf},
    labelfont=bf,
    format=hang,    
    format=plain,
    margin=0pt,
    width=0.8\textwidth,
]{caption}
\usepackage[list=true]{subcaption}
\usepackage{enumitem}
\usepackage{titlesec}
\usepackage{multirow}
\usepackage{stackrel}

\usepackage{color-edits}
\addauthor{anish}{purple}
\addauthor{dennis}{blue}
\addauthor{dev}{blue}

\usepackage[width=0.9\textwidth]{caption}

\theoremstyle{plain}

\newtheorem{lemma}{Lemma}

\newtheorem{theorem}{Theorem}
\newtheorem{proposition}{Proposition}
\newtheorem{cor}{Corollary}
\newtheorem{assumption}{Assumption}
\newtheorem{remark}{Remark}

\usepackage{prettyref}

\newrefformat{model}{Model\,\ref{#1}}
\newrefformat{listing}{Listing\,\ref{#1}}
\newrefformat{algm}{Algorithm\,\ref{#1}}
\newrefformat{line}{line\,\ref{#1}}
\newrefformat{sec}{Section\,\ref{#1}}
\newrefformat{subsec}{Subsection\,\ref{#1}}
\newrefformat{section}{Section\,\ref{#1}}
\newrefformat{appendix}{Appendix\,\ref{#1}}
\newrefformat{app}{Appendix\,\ref{#1}}
\newrefformat{def}{Definition\,\ref{#1}}
\newrefformat{defn}{Definition\,\ref{#1}}
\newrefformat{theorem}{Theorem\,\ref{#1}}
\newrefformat{ax}{\ref{#1}}
\newrefformat{proposition}{Proposition\,\ref{#1}}
\newrefformat{lemma}{Lemma\,\ref{#1}}
\newrefformat{cor}{Corollary\,\ref{#1}}
\newrefformat{corollary}{Corollary\,\ref{#1}}
\newrefformat{ex}{Example\,\ref{#1}}
\newrefformat{tab}{Table\,\ref{#1}}
\newrefformat{fig}{Fig.\,\ref{#1}}
\newrefformat{eqn}{Equation~(\ref{#1})}
\newrefformat{problem}{Problem\,\ref{#1}}
\newrefformat{assumption}{Assumption\,\ref{#1}}

\usetikzlibrary{shapes,decorations,arrows,calc,arrows.meta,fit,positioning}
\tikzset{
    -Latex,auto,node distance =1 cm and 1 cm,semithick,
    state/.style ={ellipse, draw, minimum width = 0.7 cm},
    point/.style = {circle, draw, inner sep=0.04cm,fill,node contents={}},
    bidirected/.style={Latex-Latex,dashed},
    el/.style = {inner sep=2pt, align=left, sloped}
}

\DeclareRobustCommand{\Rb}{\mathbb{R}}
\DeclareRobustCommand{\bD}{\boldsymbol{D}}

\DeclareRobustCommand{\bA}{\boldsymbol{A}}
\DeclareRobustCommand{\bH}{\boldsymbol{H}}
\DeclareRobustCommand{\bE}{\boldsymbol{E}}
\DeclareRobustCommand{\bU}{\boldsymbol{U}}
\DeclareRobustCommand{\bV}{\boldsymbol{V}}
\DeclareRobustCommand{\bS}{\boldsymbol{S}}
\DeclareRobustCommand{\bM}{\boldsymbol{M}}
\DeclareRobustCommand{\bQ}{\boldsymbol{Q}}
\DeclareRobustCommand{\bP}{\boldsymbol{P}}
\DeclareRobustCommand{\bX}{\boldsymbol{X}}

\DeclareRobustCommand{\bB}{\boldsymbol{B}}

\newcommand{\bY}{\boldsymbol{Y}}

\DeclareRobustCommand{\hA}{\widehat{{A}}}

\DeclareRobustCommand{\bhA}{\widehat{\boldsymbol{A}}}

\DeclareRobustCommand{\Ic}{\mathcal{I}}
\DeclareRobustCommand{\Ec}{\mathcal{E}}
\DeclareRobustCommand{\Pc}{\mathcal{P}}

\DeclareRobustCommand{\Oc}{\mathcal{O}}
\newcommand{\Gc}{\mathcal{G}}

\newcommand{\Vc}{\mathcal{V}}

\DeclareRobustCommand{\BCc}{\mathcal{BC}}

\DeclareRobustCommand{\Rb}{\mathbb R}
\DeclareRobustCommand{\Nb}{\mathbb N}
\DeclareRobustCommand{\Pb}{\mathbb P}

\DeclareMathOperator*{\argmax}{\arg\!\max}

\newcommand{\tw}{\widetilde{w}}
\newcommand{\tY}{\widetilde{Y}}

\newcommand{\btY}{\widetilde{\bY}}

\newcommand{\hw}{\widehat{w}}
\newcommand{\hp}{\widehat{p}}

\newcommand{\halpha}{\widehat{\alpha}}

\newcommand{\distas}[1]{\mathbin{\overset{#1}{\kern\z@\sim}}}%
\newsavebox{\mybox}\newsavebox{\mysim}
\newcommand{\distras}[1]{%
  \savebox{\mybox}{\hbox{\kern3pt$\scriptstyle#1$\kern3pt}}%
  \savebox{\mysim}{\hbox{$\sim$}}%
  \mathbin{\overset{#1}{\kern\z@\resizebox{\wd\mybox}{\ht\mysim}{$\sim$}}}%
}

\newcommand{\tbeta}{\widetilde{\beta}}

\newcommand{\hbeta}{\widehat{\beta}}

\newcommand{\Ex}{\mathbb{E}}

\newcommand\independent{\protect\mathpalette{\protect\independenT}{\perp}}
\def\independenT#1#2{\mathrel{\rlap{$#1#2$}\mkern2mu{#1#2}}}
\newcommand{\notindep}{\not\perp\!\!\!\perp}

\newcommand{\htau}{\widehat{\tau}}
\newcommand{\hu}{\widehat{u}}
\newcommand{\hv}{\widehat{v}} 

\newcommand{\AR}{\textsf{AR}} 
\newcommand{\AC}{\textsf{AC}} 
\newcommand{\eAR}{\emph{\textsf{AR}}}
\newcommand{\eAC}{\emph{\textsf{AC}}} 
\newcommand{\NR}{\textsf{NR}} 
\newcommand{\NC}{\textsf{NC}} 
\newcommand{\SI}{\textsf{SI}}
\newcommand{\USVT}{\texttt{USVT}}
\newcommand{\SNN}{\texttt{SNN}}

\newcommand{\anchor}{\texttt{AnchorSubMatrix}}
\newcommand{\softimpute}{\texttt{softImpute}} 
\newcommand{\lrsoftimpute}{\texttt{LR-softImpute}} 
\newcommand{\bitsoftimpute}{\texttt{1bitMC-softImpute}} 
\newcommand{\mmmf}{\texttt{MMMF}} 
\newcommand{\svd}{\texttt{SVD}} 
\newcommand{\svdpp}{\texttt{SVD++}} 
\newcommand{\pmf}{\texttt{PMF}}
\newcommand{\lrpmf}{\texttt{LR-PMF}}
\newcommand{\bitpmf}{\texttt{1bitMC-PMF}}
\newcommand{\lrsvd}{\texttt{LR-SVD}}
\newcommand{\bitsvd}{\texttt{1bitMC-SVD}}
\newcommand{\lrsvdpp}{\texttt{LR-SVD++}}
\newcommand{\bitsvdpp}{\texttt{1bitMC-SVD++}}
\newcommand{\maxnorm}{\texttt{MaxNorm}}
\newcommand{\lrmaxnorm}{\texttt{LR-MaxNorm}}
\newcommand{\bitmaxnorm}{\texttt{1bitMC-MaxNorm}}
\newcommand{\wtn}{\texttt{WTN}}
\newcommand{\lrwtn}{\texttt{LR-WTN}}
\newcommand{\bitwtn}{\texttt{1bitMC-WTN}}
\newcommand{\knn}{\texttt{KNN}}
\newcommand{\expomf}{\texttt{ExpoMF}}
\newcommand{\core}{\text{core}}
\newcommand{\tuser}{\text{user}}
\newcommand{\titem}{\text{item}}
\newcommand{\standard}{\text{standard}}
\newcommand{\Cc}{\mathcal{C}}

\title{Causal Matrix Completion
}
\author{Anish Agarwal, Munther Dahleh, Devavrat Shah, and Dennis Shen}
\date{}

\begin{document}

\maketitle

\begin{abstract}
Matrix completion is the study of recovering an underlying matrix from a sparse subset of noisy observations.
Traditionally, it is assumed that the entries of the matrix are ``missing completely at random'' (MCAR), i.e., 
each entry is revealed at random, independent of everything else, with uniform probability. 
This is likely unrealistic due to the presence of ``latent confounders'', i.e., unobserved factors that determine both the entries of the underlying matrix and the missingness pattern in the observed matrix.
For example, in the context of movie recommender systems---a canonical application for matrix completion---a user who vehemently dislikes horror films is unlikely to ever watch horror films.
In general, these confounders yield ``missing not at random'' (MNAR) data, which can severely impact any inference procedure that does not correct for this bias. 

We develop a formal causal model for matrix completion through the language of potential outcomes, and provide novel identification arguments for a variety of causal estimands of interest. 
We design a procedure, which we call ``synthetic nearest neighbors'' (\SNN), to estimate these causal estimands.
We prove finite-sample consistency and asymptotic normality of our estimator. 
Our analysis also leads to new theoretical results for the matrix completion literature. 
In particular, we establish entry-wise, i.e., max-norm, finite-sample consistency and asymptotic normality results for matrix completion with MNAR data. 
 As a special case, this also provides entry-wise bounds for matrix completion with MCAR data. 
Across simulated and real data, we demonstrate the efficacy of our proposed estimator. 
\end{abstract}

\newpage
\tableofcontents
\newpage 

% section 1: intro 
\section{Introduction}\label{sec:intro}

Matrix completion is the study of recovering an underlying matrix from its noisy and partial observations. 
Given its widespread applicability, the field of matrix completion has grown tremendously in recent years. 
To establish statistical guarantees for the various algorithms that exist for matrix completion, it is typically assumed that: 
(i) the underlying noiseless matrix has latent structure, e.g., it is low-rank, 
and (ii) the entries of this matrix are missing completely at random (MCAR), i.e., an entry is missing independent of everything else and with uniform probability. 
However, numerous modern applications of interest violate the latter assumption. 
Below, we consider two motivating examples. 

First, arguably the most well-known application of matrix completion is recommender systems, which are ubiquitous in modern online platforms.
Typically, data is collected in the form of a matrix, where the rows index users and columns index items; 
the $(i,j)$-th entry, therefore, corresponds to the rating supplied by user $i$ for item $j$. 
In such scenarios, observations are often subject to {\em selection-biases}. 
For instance, in movie recommendations, a fan of fantasy fiction will almost certainly watch and highly rate the {\em Harry Potter} series. 
Similarly, in restaurant recommendations, a vegetarian is unlikely to enjoy nor rate a steakhouse restaurant. 
While these examples demonstrate self-selection biases from the end of the users, systems also exhibit targeted suggestions. 
For example, when a user searches for trails at the Grand Canyon, an ad placement system is more likely to display an ad for hiking boots than wedding shoes; 
in turn, this can increase the user's likelihood to purchase and rate hiking boots. 
In all of these cases, the user's preferences and/or the system's beliefs in its users' preferences, influence the sparsity pattern of the observation matrix.

A second example is panel data settings in econometrics.
Here, observations of units (e.g., individuals, geographic locations) are collected over time as they undergo different interventions (e.g., promotions, socio-economic policies). 
The induced matrix has rows index units and columns index time-intervention pairs; 
the $(i, (a, t))$-th entry then corresponds to the potential outcome of unit $i$ under the $a$-th intervention at time step $t$; here $(a, t)$ represents the $j$-th column, i.e., columns are double indexed by both intervention and time. 
As with recommender systems, observations in panel data settings are unlikely to occur completely at random.
For instance, policy-makers strategically recommend programs that are designed to achieve certain desirable outcomes based on numerous socio-economic factors surrounding the geographic region under their purview. 
Further, competing programs with disagreeing agendas cannot be simultaneously adopted for a specific region during the same time period, i.e., if the $(i, (a, t))$-th entry is observed, then the $(i, (a', t))$-th entry must be missing.
Notably, similar matrices and observation patterns can arise in sequential decision-making paradigms within machine learning such as online learning, contextual bandits, and reinforcement learning with time-intervention pairs being replaced by state-action pairs. 

In both examples, the missingness pattern of the matrix is dependent on the underlying values in that matrix, and observing the outcome of one entry can alter the probability of observing another.
That is, the entries are missing {\em not} at random (MNAR).
To address the above challenges, there has been exciting recent progress on matrix completion with MNAR data, including \cite{schnabelfwang16, ma2019missing, zhu2019high, sportisse2020imputation_low_rank, sportisse2020estimation_PCA, causal_recommender_systems, yang2021tenips, bhattacharya2021matrix}.
% which allow for correlation between the entries of the matrix and the corresponding probabilities of observation. 
%
Through numerous empirical studies, these works have shown that algorithms that account for MNAR data outperform conventional algorithms that are designed for MCAR data.
%
% when datasets are biased, which is often the case in real-word settings. 
%
%MNAR matrix completion algorithms reduce the biases that emerge from  
%
With respect to theoretical analysis, however, critical aspects of matrix completion with MNAR data remain to be explored. 
In particular, as highlighted in \cite{ma2019missing}, there are two common limiting assumptions in the literature: 
(i) the revelation of each entry in the matrix is independent of all other entries, 
and 
(ii) each entry has a nonzero probability of being observed. 
Another recent exciting line of work that we build upon is that of panel data and matrix completion, see \cite{RSC, mRSC, arkhangelsky2019synthetic, bai2019matrix, fernandez2020low, athey2021matrix,  agarwal2019robustness,agarwal2020principal, agarwal2021synthetic, agarwal2021causal}.
Some of these works allow for MNAR data and entries of a matrix to be deterministically missing.
However, they consider very restricted sparsity patterns that are not particularly suitable for important applications of matrix completion.
For example, the most common sparsity pattern considered in the panel data literature is where for a given row $i$, if a column $j$ is missing, then entries for all columns $j' > j$ in row $i$ are also missing; such a pattern is unlikely to arise in recommendation systems or sequential decision-making.
Further, to the best of our knowledge, none of these works within the panel data literature provide meaningful results for matrix completion with MCAR data.
The statistical parameters these works aim to estimate are also less meaningful for these other applications of matrix completion.
The most common statistical parameter these works consider is the average outcome for all missing entries in a given row $i$; in say recommendation systems, this would correspond to the average rating a user $i$ would have given for all movies they did not rate.
This is not particularly meaningful for an online platform---ideally, a platform would like to do accurate inference for each $(i, j)$ pair. 
The focus of this work is to propose a formal causal framework and an algorithm with provable guarantees to analyze matrix completion with MNAR data where the probability that an entry of the matrix is missing can: (i) depend on the underlying values in the matrix itself; (ii) depend on which other entries are missing; (iii) potentially be deterministically zero.
Further, we want to allow for more general missingness patterns and estimate more refined statistical parameters than considered in the panel data literature thus far. 

Indeed, it is both commonly said that
%
% \vspace{2pt}
\begin{center}
\textit{``Causal inference is a missing data problem.''} \\
{\small \&}\\
\textit{``Matrix completion is a missing data problem.''}
\end{center}
% \vspace{2pt}
%
We hope this work further bridges the rich and growing fields of causal inference and matrix completion.

\subsection{How the Missingness Mechanism can Bias Inference: A Teaser}\label{sec:intro_teaser}
\begin{figure}[h!]
	\centering 
	\begin{subfigure}[b]{0.2\textwidth}
		\centering 
		\includegraphics[width=\linewidth]
		{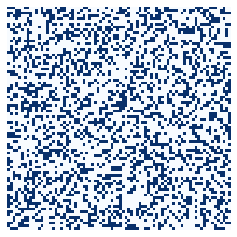}
		\caption{\smaller MCAR.} 
		\label{fig:sparsity_MCAR}
	\end{subfigure} 
	\qquad	\begin{subfigure}[b]{0.2\textwidth}
		\centering 
		\includegraphics[width=\linewidth]
		{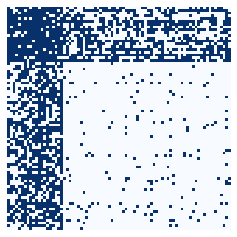}
		\caption{\smaller Limited MNAR.} 
		\label{fig:sparsity_limited_MNAR} 
	\end{subfigure}
	\qquad 
	\begin{subfigure}[b]{0.2\textwidth}
		\centering 
		\includegraphics[width=\linewidth]
		{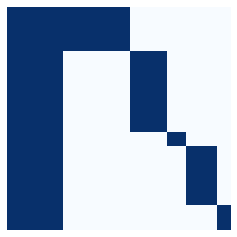}
		\caption{\smaller General MNAR.} 
		\label{fig:sparsity_general_MNAR} 
	\end{subfigure} 
	\caption{\smaller Empirical sparsity pattern under different missingess mechanisms.}
	\label{fig:sparsity_patterns} 
\end{figure}

As further motivation for why it is important to carefully think about the underlying mechanism for why data is missing, we now provide illustrative empirical simulations.
In particular we run three experiments, each with a different mechanism for how data is missing.
In Experiment 1, data is missing via a MCAR mechanism i.e., each entry is missing independently at random with probability $0.35$; the induced sparsity pattern is depicted in Figure \ref{fig:sparsity_MCAR}.
In Experiment 2, data is missing in a MNAR fashion, i.e., each entry has a different probability of being missing; the induced sparsity pattern is depicted in Figure \ref{fig:sparsity_limited_MNAR}.
However, we ensure key assumptions made thus far in the matrix completion literature with MNAR data are maintained; in particular, (i) the revelation of entries are entry-wise independent and (ii) each entry has a nonzero probability of being observed. 
In Experiment 3, data is missing in a MNAR fashion, but we violate conditions (i) and (ii) above; the induced sparsity pattern is depicted in Figure \ref{fig:sparsity_general_MNAR}.
For exact details on the missingness mechanism in Experiment 2 and 3, refer to Section \ref{sec:standardmnar} and \ref{sec:generalmnar}, respectively.

\begin{figure}[h!]
	\centering 
	\begin{subfigure}[b]{0.24\textwidth}
		\centering 
		\includegraphics[width=\linewidth]
		{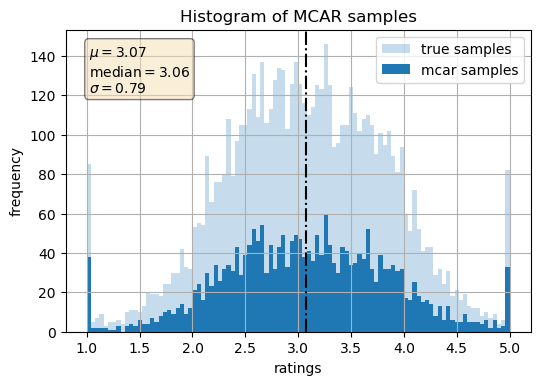}
		\caption{\smaller True and revealed ratings.} 
		\label{fig:teaser_MCAR_true}
	\end{subfigure} 
	\hfill	\begin{subfigure}[b]{0.24\textwidth}
		\centering 
		\includegraphics[width=\linewidth]
		{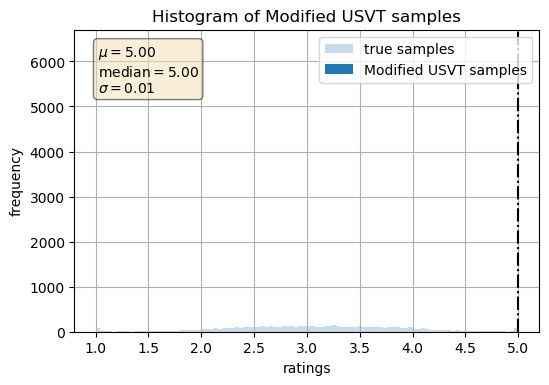}
		\caption{\smaller (Modified) \USVT.} 
		\label{fig:teaser_MCAR_USVT} 
	\end{subfigure}
	\hfill 
	\begin{subfigure}[b]{0.24\textwidth}
		\centering 
		\includegraphics[width=\linewidth]
		{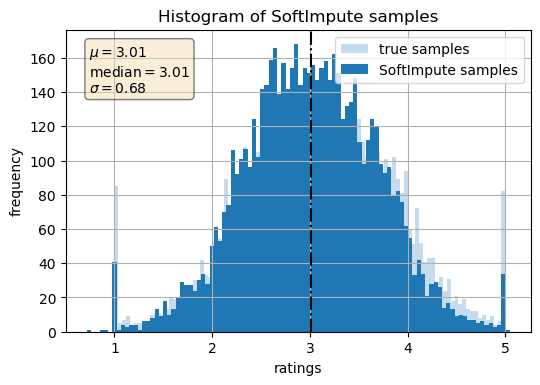}
		\caption{\smaller (Modified) \softimpute.} 
		\label{fig:teaser_MCAR_softimpute} 
	\end{subfigure} 
	\begin{subfigure}[b]{0.24\textwidth}
		\centering 
		\includegraphics[width=\linewidth]
		{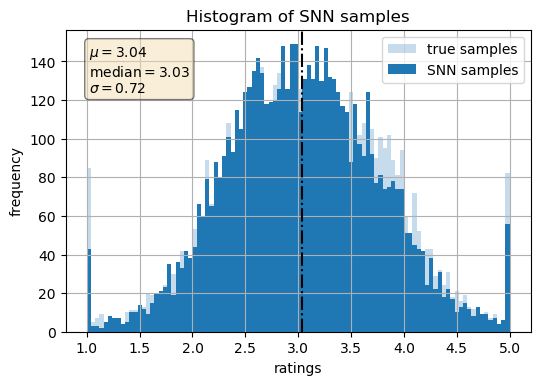}
		\caption{\smaller \SNN.} 
		\label{fig:teaser_MCAR_SNN} 
	\end{subfigure}
	\caption{\smaller MCAR: recovered ratings distributions under (modified) \USVT, (modified) \softimpute, and \SNN.}
	\label{fig:sparsity_patterns_MCAR} 
\end{figure}

\begin{figure}[h!]
	\centering 
	\begin{subfigure}[b]{0.24\textwidth}
		\centering 
		\includegraphics[width=\linewidth]
		{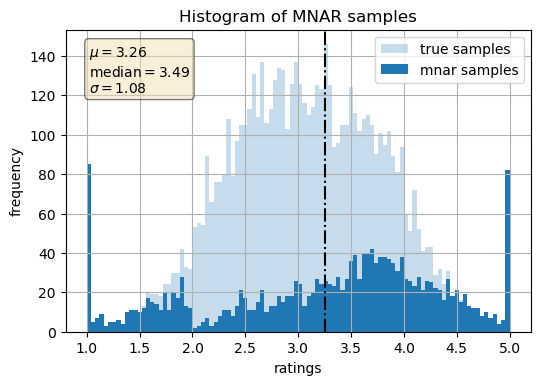}
		\caption{\smaller True and revealed ratings.} 
		\label{fig:teaser_limited_MNAR_true}
	\end{subfigure} 
	\hfill	\begin{subfigure}[b]{0.24\textwidth}
		\centering 
		\includegraphics[width=\linewidth]
		{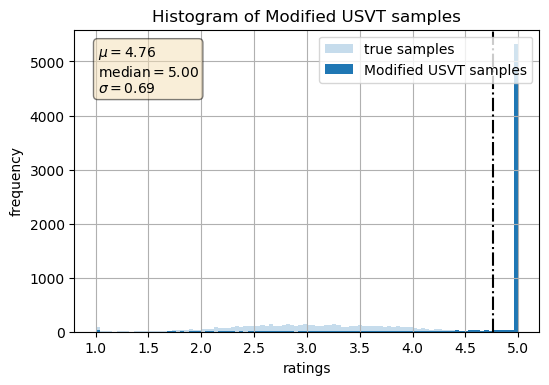}
		\caption{\smaller (Modified) \USVT.} 
		\label{fig:teaser_limited_MNAR_USVT} 
	\end{subfigure}
	\hfill 
	\begin{subfigure}[b]{0.24\textwidth}
		\centering 
		\includegraphics[width=\linewidth]
		{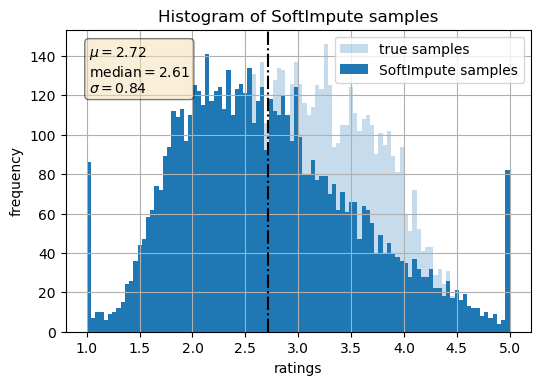}
		\caption{\smaller (Modified) \softimpute.} 
		\label{fig:teaser_limited_MNAR_softimpute} 
	\end{subfigure} 
	\begin{subfigure}[b]{0.24\textwidth}
		\centering 
		\includegraphics[width=\linewidth]
		{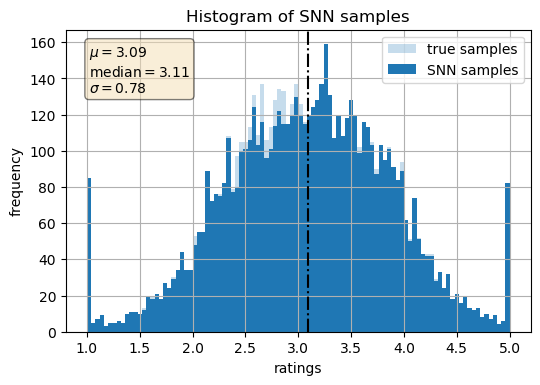}
		\caption{\smaller \SNN.} 
		\label{fig:teaser_MCAR_SNN} 
	\end{subfigure}
	\caption{\smaller Limited MNAR: recovered ratings distributions under (modified) \USVT, (modified) \softimpute, and \SNN.}
	\label{fig:sparsity_patterns_limited_MNAR} 
\end{figure}

\begin{figure}[h!]
	\centering 
	\begin{subfigure}[b]{0.24\textwidth}
		\centering 
		\includegraphics[width=\linewidth]
		{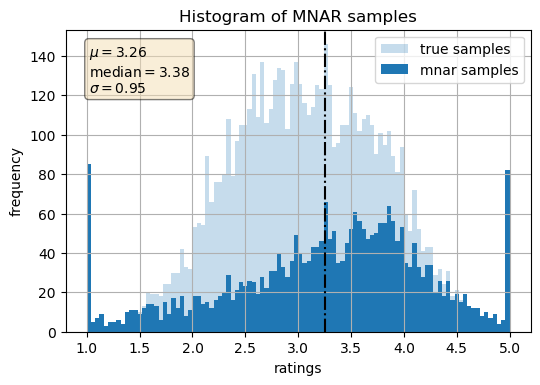}
		\caption{\smaller True and revealed ratings.} 
		\label{fig:teaser_general_MNAR_true}
	\end{subfigure} 
	\hfill	\begin{subfigure}[b]{0.24\textwidth}
		\centering 
		\includegraphics[width=\linewidth]
		{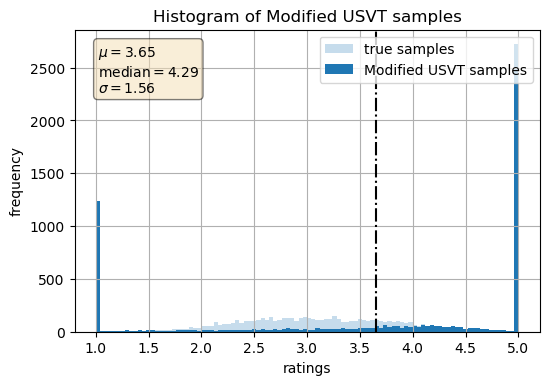}
		\caption{\smaller (Modified) \USVT.} 
		\label{fig:teaser_general_MNAR_USVT} 
	\end{subfigure}
	\hfill 
	\begin{subfigure}[b]{0.24\textwidth}
		\centering 
		\includegraphics[width=\linewidth]
		{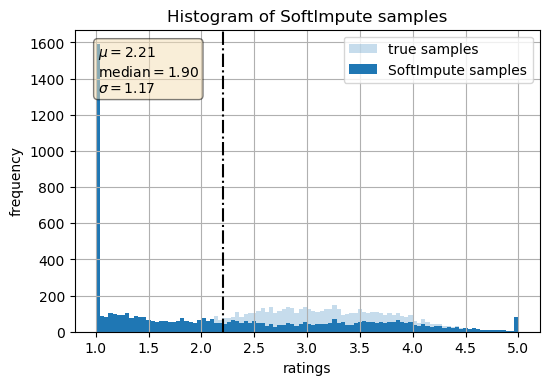}
		\caption{\smaller (Modified) \softimpute.} 
		\label{fig:teaser_general_MNAR_softimpute} 
	\end{subfigure} 
	\begin{subfigure}[b]{0.24\textwidth}
		\centering 
		\includegraphics[width=\linewidth]
		{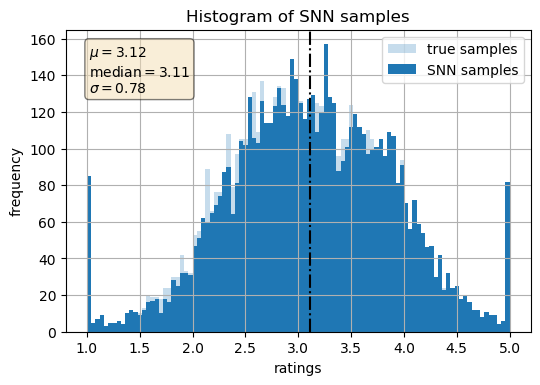}
		\caption{\smaller \SNN.} 
		\label{fig:teaser_general_MNAR_SNN} 
	\end{subfigure}
	\caption{\smaller More general MNAR: recovered ratings distributions under (modified) \USVT, (modified) \softimpute, and \SNN.}
	\label{fig:sparsity_patterns_general_MNAR} 
\end{figure}

In all experiments, we first create a sample of true ``ratings'', which are invariant across all three experiments.
We enforce these ratings to go from 1 to 5, as is standard in many online platforms.
The distribution of true/revealed ratings are plotted in light/dark blue in Figures \ref{fig:teaser_MCAR_true}, \ref{fig:teaser_limited_MNAR_true}, and  \ref{fig:teaser_general_MNAR_true}, respectively.
As expected, the distribution of the revealed ratings in the MCAR setup matches that of the true ratings.
However, the set of ratings that are revealed in both MNAR settings are severely biased, i.e., their distribution does not match that of the true underlying ratings.

We use three matrix completion algorithms and see whether they can recover the distribution of true ratings given the revelead entries in all three experiments.
The algorithms are:
Universal singular value thresholding (\USVT)  \cite{Chatterjee15}, which is a popular spectral based method;\footnote{Surprisingly, we find that the original \USVT~algorithm performs better in all three experiments. See Appendix \ref{teaser:standard_USVT}.}
Softimpute (\softimpute) \cite{softimpute14}, which is a popular optimization based method;
Synthetic nearest neighbours (\SNN), which is our proposed method for matrix completion with MNAR data, and is a combination of the approach taken in nearest neighbour style and panel data methods in econometrics.
\USVT~and \softimpute~are not designed for MNAR data, as is, but we de-bias them for MNAR data as is done in \cite{bhattacharya2021matrix} and \cite{ma2019missing}, respectively.
See details in Section \ref{sec:empirics}.

We see that in Figure \ref{fig:sparsity_patterns_MCAR}, under the MCAR setting, \softimpute~and \SNN~both recover the distribution of true ratings very well, while  \USVT~cannot.
Once we go to the limited MNAR setting, depicted in Figure \ref{fig:sparsity_patterns_limited_MNAR}, where conditions (i) and (ii) are upheld, \SNN~is still able to recover the underlying distribution of true ratings, but now both \softimpute~and \USVT~have non-negligible bias.
In the general MNAR setting, depicted in Figure \ref{fig:sparsity_patterns_general_MNAR}, \SNN~continues to accurately recover the distribution, but the bias of \softimpute~is significantly worsened.

This empirical illustration highlights the sensitivity of these traditional matrix completion methods to the missingness mechanism and strongly motivates the need for a rigorous framework for tackling the general MNAR setting where conditions (i) and (ii) above are violated.
Providing such a framework is what we set out to do in this work.

\subsection{Problem Statement}\label{sec:problem_setup}
We now formally introduce our setup.
Consider a signal matrix $\bA = [A_{ij}] \in \Rb^{m \times n}$, 
a noise matrix $\bE = [\varepsilon_{ij}] \in \Rb^{m \times n}$,
and a propensity score matrix $\bP = [p_{ij}] \in [0,1]^{m \times n}$. 
All three matrices are entirely latent, i.e.,  unobserved. 
Let $\bY = [Y_{ij}] \in \Rb^{m \times n}$ denote the ``noisy'' version of $\bA$, with $\Ex[\bY] = \bA$; we denote $\varepsilon_{ij} = Y_{ij} - A_{ij}$.
We assume $\bY$ itself is partially observed. 
In particular, we denote $\bD = [D_{ij}] \in \{0,1\}^{m \times n}$ with $\Ex[\bD] = \bP$ as the missingness mask matrix that indicates which entries of $\bY$ are observed. 
For convenience, we encode our observations into $\btY = [\tY_{ij}] \in \{\Rb \cup \{\star\}\}^{m \times n}$ such that for $(i,j) \in [m] \times [n]$,
\begin{align}\label{eq:problem_setup}
   \tY_{ij} &= \begin{cases}
        &Y_{ij}, \text{ if } D_{ij} = 1
        \\
        & \star, \text{ otherwise.} 
    \end{cases}
\end{align} 
In words, if $D_{ij}=1$ then $A_{ij}$ is noisily observed, and if $D_{ij}=0$ then $A_{ij}$ remains unknown. 
For concreteness, let us return to the recommender system example. 
Here, $\bA$ represents the expected rating for every user-item pair and $\bP$ dictates the probability these expected ratings are revealed, both of which are unknown. 
$\bY$ in relation to $\bA$ then models the inherent randomness in how users rate items; that is, $\bY$ can be interpreted as a ``noisy'' instance of $\bA$. 
Another interpretation of what $\varepsilon_{ij}$ represents is that many online platform only allow users to input integer valued ratings (e.g. integer between 1 to 5 or a binary 0/1).
Hence, $Y_{ij}$ can be interpreted as a ``noisy'' {\em discretized} observation of $A_{ij}$, which may actually be {\em continuous} (i.e., lie within the continuous interval $[1,5]$ or $[0, 1]$). 
Observationally, we have access to $\bD$ and $\btY$; the former refers to the collection of ratings users have supplied to the system while the latter refers to the corresponding realized ``noisy'' ratings. 
Finally, we remark that \eqref{eq:problem_setup} also agrees with standard panel data setups in econometrics, where each observation is assumed to be corrupted by an idiosyncratic shock, which is represented by $\varepsilon_{ij}$.  

In terms of the type of MNAR data this work considers, we allow for $\bD$ and $\bY$ to be dependent, provided $\bD \independent \bY | \bA$, where $\bA$ is latent.
In fact, we allow $\bD$ to be any arbitrary function of $\bA$, random or deterministic, subject to suitable observation patterns which we discuss in the forthcoming sections. 
Notably, our framework also allows the entries in $\bD$ to be dependent with each other across both rows and columns, and the minimum value of $\bP$ to be $0$, which are important departures from the current matrix completion literature. 
Under these conditions, we propose an algorithm that provably recovers $\bA$ from $\btY$ with entry-wise (i.e., max-norm) guarantees.

\subsection{Contributions \& Paper Organization}
{\bf Section~\ref{sec:related_works}: Related works.} 
We provide an overview of the current literature on matrix completion under the different models of missingness proposed by \cite{rubin1976, little2019statistical}: (i) missing completely at random (MCAR); (ii) missing at random (MAR); (iii) missing not at random (MNAR). 
We note the usage of the terms MCAR, MAR, and MNAR is inconsistent across the previous works on matrix completion, and so we hope that our literature survey helps give a more comprehensive and unified overview of the different regimes of missingness considered in these works. 
 
{\bf Section~\ref{sec:causal_framework}: Causal framework for matrix completion.} 
We propose a formal causal framework for matrix completion using the language of potential outcomes, see \cite{neyman, rubin}. 
We interpret $\bY$ as the matrix of potential outcomes
and $\bP$ as the matrix of intervention assignments. 
Building upon the recent work of \cite{agarwal2021synthetic}, we propose a framework that allows 
(i) correlation between $\bD$ and $\bY$, i.e., hidden confounding; 
(ii) correlation between the entries of $\bD$; 
(iii) the minimum value of $\bP$ to be $0$, i.e., entries of $\btY$ can be deterministically missing; 
(iv) $\bP$ to not exhibit low-dimensional structure as is required in the panel data literature, i.e., we consider significantly more general missingness patterns.
To the best of our knowledge, our framework, and associated algorithm, is the first within the MNAR matrix completion literature that allows for conditions (i)-(iv) to simultaneously hold.
Additionally, we do {\em not} make any parametric or distributional assumptions on $\bP$, as is common in previous works on matrix completion.
Nevertheless, we establish an identification result in Theorem \ref{thm:identification}, which effectively states that $\bA$ can be learned from $\btY$ in an entry-wise sense.
We believe our proposed framework provides a unified causal view for a variety of applications that can be posed as matrix completion problems with MNAR data.

{\bf Section~\ref{sec:estimator}: An algorithmic solution.} 
We combine the nearest neighbours approach for matrix completion ---popularly known as collaborative filtering---with the synthetic controls approach for panel data, to design a novel two-step algorithm, which we call ``synthetic nearest neighbors'' (\SNN), to estimate $\bA$ from $\btY$. 
Pleasingly, each step of \SNN~enjoys a simple closed-form solution. 
In order to efficiently execute \SNN~in practice, we provide an algorithm to automatically find the ``neighbors'' for any $(i,j)$ pair in a data-driven manner. 
To do so, we relate this task to the well-known problem of finding the ``maximum'' biclique in a bipartite graph. 
Since \SNN~is a generalization of the recently proposed synthetic interventions (\SI)~estimator of \cite{agarwal2021synthetic}, which itself is a generalization of the popular synthetic controls algorithm of \cite{abadie1, abadie2}, this subroutine may be of independent interest to the synthetic controls and panel data literatures.

{\bf Section~\ref{sec:theoretical_results}: Theoretical results.} 
We establish entry-wise finite-sample consistency and asymptotic normality of \SNN, i.e., we provide theoretical guarantees for $A_{ij}$ for each $(i,j)$ pair.
Hence, our analysis implies new theoretical results, in a max-norm sense, for the literature on matrix completion with MNAR data. 
As a special case, this also provides novel entry-wise finite-sample consistency and asymptotic normality results for the traditional matrix completion with MCAR data literature.
Collectively, our identification, consistency, and asymptotic normality results, coupled with \SNN, can be seen as a generalization of the \SI~framework proposed in \cite{agarwal2021synthetic}.

{\bf Section~\ref{sec:experiments}: Experimental validation.}
We run comprehensive experiments, both with simulated and real-world data, to test the empirical efficacy of \SNN~against a collection of state-of-the-art matrix completion algorithms for MNAR data.
Some key takeaways are as follows: 
(i) \SNN~is robust to the various forms of missingness across all experiments, while the previous methods are relatively sensitive to it.
(ii) we find the approaches to de-bias estimators for MNAR data are not particularly effective, i.e., their performance is similar to their MCAR analogues; this is in line with the empirical findings of \cite{ma2019missing}.
%

% NOTATIONS
\subsection{Notations} \label{sec:notations} 
For a matrix $\bX \in \Rb^{m \times n}$, we denote its operator (spectral), nuclear, Frobenius, and max element-wise norms as $\|\bX\|_{2}$, $\|\bX\|_*$, $\|\bX\|_F$, and $\|\bX\|_{\max}$, respectively.  
For a matrix $\bX$ with orthonormal columns, let $\Pc_X = \bX \bX^T$ denote the projection matrix onto the subspace spanned by the columns of $\bX$. 
For a vector $v \in \Rb^m$, let $\|v\|_p$ denote its $\ell_p$-norm.
For a random variable $v$, we define its sub-gaussian (Orlicz) norm as $\|v\|_{\psi_2}$. 
Let $\circ$ denote component-wise multiplication and let $\otimes$ denote the outer product.
For a positive integer $a$, let $[a] = \{1, \dots, a\}$. 
For index sets $\Ic_1 \subseteq [m] $ and $\Ic_2 \subseteq [n]$, let $\bX_{\Ic_1, \Ic_2}$ denote the $| \Ic_1| \times | \Ic_2|$ sub-matrix of $\bX$ whose rows and columns are indexed by $\Ic_1$ and $\Ic_2$, respectively.
As a shorthand, let $\bX_{\Ic_1, \cdot}$ denote the $|\Ic_1| \times n$ sub-matrix of $\bX$ that retains the columns of $\bX$ but only considers those rows indexed by $\Ic_1$; we define $\bX_{\cdot, \Ic_2}$ analogously. 
Unless stated otherwise, we index rows with $i \in [m]$ and columns with $j \in [n]$. 

Let $f$ and $g$ be two functions defined on the same space. 
We say $f(n)$ = $O(g(n))$ if and only if there exists a positive real number $M$ and a real number $n_0$ such that for all $n \ge n_0, |f (n)| \le M|g(n)|$.
Analogously we say:
$f (n) = \Theta(g(n))$ if and only if there exists positive real numbers $m, M$ such that for all $n \ge n_0, \ m|g(n)| \le |f(n)| \le M|g(n)|$;
$f (n) = o(g(n))$ if for any $m > 0$, there exists $n_0$ such that for all $n \ge n_0, |f(n)| \le m|g(n)|$.
We adopt the standard notations and definitions for stochastic convergences. 
As such, we denote $\xrightarrow{d}$ and $\xrightarrow{p}$ as convergences in distribution and probability, respectively. 
We will also make use of $O_p$ and $o_p$, which are probabilistic versions of the commonly used deterministic $O$ and $o$ notations.
More formally, for any sequence of random vectors $X_n$, we say $X_n = O_p(a_n)$ if for every $\varepsilon>0$, there exists constants $C_\varepsilon$ and $n_\varepsilon$ such that $\Pb( \| X_n \|_2 > C_\varepsilon a_n) < \varepsilon$ for every $n \ge n_\varepsilon$; 
equivalently, we say $(1/a_n) X_n$ is ``uniformly tight'' or ``bounded in probability''.
Similarly, $X_n = o_p(a_n)$ if for all $\varepsilon, \varepsilon'> 0$, there exists $n_\varepsilon$ such that $\Pb(\| X_n \|_2 > \varepsilon' a_n) < \varepsilon$ for every $n \ge n_\varepsilon$.
Therefore, $X_n = o_p(1) \iff X_n \xrightarrow{p} 0$. 
Additionally, we denote: $\text{plim} \ X_n = a \iff X_n \xrightarrow{p} a$. 
We say a sequence of events $\Ec_n$, indexed by $n$, holds ``with high probability'' (w.h.p.) if $\Pb(\Ec_n) \rightarrow 1$ as $n \rightarrow \infty$, i.e., for any $\varepsilon > 0$, there exists a $n_\varepsilon$ such that for all $n > n_\varepsilon$, $\Pb(\Ec_n) > 1 - \varepsilon$. More generally, a multi-indexed sequence of events $\Ec_{n_1,\dots, n_d}$, with indices $n_1,\dots, n_d$ with $d \geq 1$, is said to hold w.h.p. if $\Pb(\Ec_{n_1,\dots, n_d}) \rightarrow 1$ as $\min\{n_1,\dots, n_d\} \rightarrow \infty$.
We also use $\mathcal{N}(\mu, \sigma^2)$ to denote a normal or Gaussian distribution with mean $\mu$ and variance $\sigma^2$---we call it {\em standard} normal if $\mu = 0$ and $\sigma^2 = 1$.

% section 2: related works
\section{Related Works}\label{sec:related_works}
Given the vastness of the matrix completion literature, we do not strive to do an exhaustive review of it.
Instead, we focus on a few representative works that propose and analyze algorithms designed for the three different models of missingness: MCAR, MAR, and MNAR.
In Section~\ref{sec:me_architecture}, we give an overview of the type of algorithms for matrix completion studied thus far in existing works. 
In Section~\ref{sec:missingness_models}, we discuss the different models of missingness considered in the matrix completion literature, and representative algorithms for these various models.
Finally, in Section~\ref{sec:panel_data_review}, we discuss the growing literature exploring the intersection of matrix completion and causal inference; in particular, the panel data literature in econometrics. 

\subsection{Overview of Matrix Completion Algorithms} \label{sec:me_architecture}
Algorithms for matrix completion broadly fall into two classes: empirical risk minimization (ERM) methods and matching (i.e., collaborative filtering) methods, with ERM methods being relatively more popular.
We give an overview of both class of methods below.

{\bf Empirical Risk Minimization (ERM) Methods.}
Empirical risk minimization (ERM) is arguably the de facto approach to recover the underlying signal matrix $\bA$ given $\btY$. 
Specifically, ERM approaches aim to solve the following program: 
\begin{align} \label{eq:erm_uniform} 
\textsf{minimize} ~~\frac{1}{|\Omega|} &\sum_{(i,j) \in \Omega} d( T_{ij}, Q_{ij} ) + \lambda ~\textsf{regularize}(\bQ).
%\quad \\ &\text{s.t.} \ \textsf{constraint}(\bQ) \ \text{holds}.
\end{align}
Here, $\Omega \subseteq [m] \times [n]$, $d(\cdot, \cdot)$ is an appropriate distance measure (e.g., squared loss), 
$T_{ij}$ is a ``simple'' transformation of $\tY_{ij}$ (e.g. $\mathds{1}(D_{ij}=1) \cdot \tY_{ij}$), 
$\textsf{regularize}(\cdot)$ is a regularization term and $\lambda > 0$ is the regularization hyper-parameter. 
For certain algorithms, they  replace the regularizer (i.e., set $\lambda = 0$) with a constraint, $\textsf{constraint}(\cdot)$.  

In order to prove statistical guarantees about these various estimators, structure is placed on $\bA$. 
The assumptions made guide the specific choices of the above parameters, which then define the algorithm. 
For instance, if the singular values of $\bA$ are assumed to be moderately sparse (i.e., only few are non-zero), then a natural convex regularizer would penalize solutions with large nuclear norm, i.e., $\textsf{regularize}(\bQ) = \| \bQ \|_*$ \cite{CandesTao10, Recht11}. 
Indeed, choosing $\Omega = \{(i,j): D_{ij} = 1\}$ as the collection of observed entries, $T_{ij} = \tY_{ij}$, and $d(\cdot, \cdot)$ as the squared loss yields the popular \softimpute~algorithm of \cite{softimpute, softimpute14}. 
As another example, if $\bA$ is assumed to be exactly low-rank, then a natural constraint would be the rank of the output matrix. 
More specifically, $\textsf{contraint}(\bQ)$ can be defined as $\text{rank}(\bQ) \le \mu$ for some pre-specified integer $\mu > 0$. 
Then, choosing $\Omega = [m] \times [n]$, $T_{ij} = \mathds{1}(D_{ij}=1) \cdot \tY_{ij} $, and $d(\cdot, \cdot)$ as the squared loss yields a suite of spectral based methods \cite{KeshavanMontanariOh10a, KeshavanMontanariOh10b, donoho14, Chatterjee15}. 
Other notable algorithms within the broader ERM class include maximum-margin matrix factorization (\mmmf)~\cite{mmmf}, probabilistic matrix factorization (\pmf)~\cite{pmf}, and \svdpp~\cite{koren08} to name a few.

Broadly speaking, it is commonly assumed that $\bA$ follows some form of a latent variable model; in particular, $A_{ij} = f(u_i, v_j)$, where $f$ is a sufficiently ``smooth'' latent function (e.g., H\"older continuous), and $u_i, v_j$ are low-dimensional latent variables associated with row $i$ and column $j$, respectively.
Such latent variable models imply that $\bA$ is (approximately) low-rank, i.e., $A_{ij} \approx \langle u_i, v_j \rangle$, where $u_i, v_j \in \Rb^r$ and $r \ll \min\{m, n\}$, e.g., \cite{xu2017rates, udell_low_rank, agarwal2019robustness}.
For an excellent overview on standard assumptions made on $\bA$ and the subsequent guarantees proven for the estimation error, please refer to \cite{ieee_matrix_completion_overview}.

When every entry is revealed with uniform probability (i.e., $p_{ij} = p$), \eqref{eq:erm_uniform} is an unbiased estimate of the full loss function with all entries revealed (i.e., $\bD$ is an all ones matrix). 
When $p_{ij}$ are nonuniform, however, recent works have provably and empirically shown that \eqref{eq:erm_uniform} is biased \cite{schnabelfwang16, ma2019missing}.
As such, these works advocate to de-bias the standard ERM objective by re-weighting each observation inversely by its propensity score $p_{ij}$.
This technique is often known in the causal inference literature as inverse propensity scoring (IPS) or weighting (IPW), see \cite{imbens_rubin_2015, little2019statistical}. 
This yields the following adapted program: 
\begin{align} \label{eq:erm_weighted}
\textsf{minimize} &\sum_{(i,j) \in \Omega} (1/\hp_{ij}) ~ d( T_{ij}, Q_{ij} ) + \lambda ~\textsf{regularize}(\bQ),
%\\ &\text{s.t.} \ \textsf{constraint}(\bQ) \ \text{holds}.
\end{align}
where $\hp_{ij}$ is an estimate of $p_{ij}$. 
In words, \eqref{eq:erm_weighted} requires learning $\bP$ prior to carrying out the standard ERM of \eqref{eq:erm_uniform}. 
Faithful matrix recovery under more general missingness patterns thus requires structure on not only $\bA$, but also $\bP$ and $\bD$. 
We overview standard assumptions on these quantities in Section \ref{sec:missingness_models}. 

{\bf Matching methods.}
For traditional applications of matrix completion, such as recommendation systems, K nearest neighbour (\knn) methods have been popular (e.g., \cite{goldberg1992using, linden2003amazon, kleinberg2008using, koren2015advances, LeeLiShahSong16, chen2018explaining}).
In \knn, to impute a missing entry $(i, j)$, the first step is to select $K$ rows for which the entry in the $j$-th column is not missing.
Of all the rows for which the $j$-th column is not missing, the $K$ rows are selected such that they are the ``closest'' to row $i$.
In particular, a hyper-parameter of \knn~is the metric that is chosen to define ``closeness'' between any two given rows; the most commonly used metric is the mean squared distance between the commonly revealed entries for a given two rows.
Once these $K$ ``neighbour rows'' are chosen, the estimate for the missing entry $(i, j)$ is the average  $\frac{1}{K}\sum_{k \in \text{neighbour rows}} \tY_{k j}$.
An attractive quality of these \knn~methods is that they do not require imputing missing values by $0$.
A related literature that shares similarities with \knn~is that of synthetic controls \cite{abadie1, abadie2}.
A key difference is that to impute $(i, j)$, uniform weights (i.e., $1/K$) are not used for the neighbouring rows; classically in synthetic controls, these weights are constrained to lie within the simplex, i.e., the weights are non-negative and sum to $1$ (if the weights are restricted to be $1/K$, this is known in the panel data literature as ``difference-in-differences'').
However, as discussed earlier, synthetic controls methods have been designed to handle restricted sparsity patterns naturally arising in 
the panel data setting. %compared to \knn~methods.
Given the growing literature on synthetic controls, we do a detailed literature review of it in Section \ref{sec:panel_data_review}.

\subsection{Three Models of Missingness}\label{sec:missingness_models}
Below, we utilize the useful taxonomy set in \cite{rubin1976, little2019statistical} to discuss the three primary mechanisms that lead to missing data and how previous works fit within these regimes.

{\bf Missing completely at random (MCAR).}
MCAR is the most standard model of missingness assumed in the matrix completion literature and is  characterized by the following properties: 
(i) $\bD \independent \bY$;
(ii) $D_{ij} \independent D_{ab}$ for all $(i,j) \neq (a,b)$;
(iii) $p_{ij} = p > 0$ for all $(i,j)$. 
In words, MCAR assumes each element of $\bD$ is an independent and identically distributed (i.i.d.) Bernoulli random variable (r.v.) with parameter $p \in (0, 1]$. 
This implies that the missingness pattern is independent of the values in $\bY$.
We note that this condition $p_{ij} > 0$ is known in the causal inference literature as ``positivity'', see \cite{imbens_rubin_2015}.
It follows that the maximum likelihood estimator $\hp_{ij} = \hp$ for all $(i,j)$, where $\hp$ is the fraction of observed entries in $\btY$. 
As previously mentioned, given MCAR data, \eqref{eq:erm_uniform} is an unbiased estimator of the ideal loss function where all entries observed. 
Though MCAR is likely unrealistic outside experimental settings, the MCAR regime remains a popular abstraction in machine learning and statistics to study the inherent trade-offs between the observation probability $p$, properties of the noise $\bE$, and the structure imposed on the signal $\bA$, in terms of the estimation error between $\bhA$ and $\bA$. 
%
% Some additional representative algorithms to the ones stated in Section~\ref{sec:me_architecture} include nuclear norm minimization and nearest neighbor methods, e.g., \cite{CandesTao10, Recht11, LeeLiShahSong16}.
%
Methods such as singular value thresholding explicitly impute missing values in $\bY$ (denoted as $\star$) by $0$ and re-weight all non-missing values in $\bY$ by $1 / \hp$, where $\hp$ is the fraction of observed entries.
This can be interpreted as a form of uniform IPW.
Other methods such as nuclear norm minimization, alternating least squares, and nearest neighbour methods do not require imputing missing values by $0$.
However, existing theoretical analysis of these algorithms do still require that $\Ex[D_{ij}] = p$, and that $D_{ij}$ is independent of the all other randomness in the model.

{\bf Missing at random (MAR).}
\footnote{Many works in the matrix completion literature do not differentiate between MAR and MNAR, and call both regimes MNAR. We differentiate between them to be more in line with models of missingness proposed by \cite{rubin1976, little2019statistical}.}
MAR is a more challenging setting than MCAR. 
The three key assumptions of MAR are as follows.
(i) $\bD \independent \bY ~|~ \Oc$, where $\Oc$ represents observed covariates about the rows and columns of the matrix (e.g., covariates about users and movies in the context of recommender systems)---concretely, these observed variables, $\Oc$, often include features or covariates $(X_i, \tilde{X}_j)$, which are associated with row $i$ and column $j$, respectively, and observed outcomes $\tY_{ij}$.
(ii) $D_{ij} \independent D_{ab}$ for all $(i,j) \neq (a,b)$.
(iii) $p_{ij} > 0$ for all $(i,j)$.
Here, the entries of $\bD$ continue to obey positivity and remain independent Bernoulli r.v.'s. 
%
% {\color{red}
% However, they are no longer necessarily identically distributed, and the differences in the distribution of missing and observed data can be  accounted for by a set of observed variables $\Oc$, i.e., the missingness pattern only depends on observed quantities.}
%
 
%
Below, we overview two popular propensity estimation techniques of \cite{schnabelfwang16}. 
To aid the following discussion, let $\bX = \{(X_i, \tilde{X}_j): (i, j) \in [m] \times [n]\}$ denote the set of observed features, and $\bH$ denote the set of hidden features. 
% $\bU = \{U_i: i \in [N]\}$, $\bV = \{V_j: j \in [M]\}$ denote the set of all latent features. 
%
The first approach is via Naive Bayes, which assumes that 
$p_{ij} = \Ex[D_{ij} | \bX, \bH, \bY] = \Ex[D_{ij} | \tY_{ij}]$. 
Under this assumption, the maximum likelihood estimator $\hp_{ij}$ can be solved using Bayes formula; 
however, such an approach requires a small sample of MCAR data, see \cite{schnabelfwang16}. 
The second estimation strategy is based on logistic regression. 
Here, it is assumed that there exists model parameters $\phi$ such that 
$p_{ij} = \Ex[D_{ij} | \bX, \bH, \bY] = \Ex[D_{ij} | X_i, \tilde{X}_j, \phi]$; 
within the causal inference literature, this is often known as ``selection on observables'', see \cite{imbens_rubin_2015}.
Typically, it is posited that  $\phi = (\omega_1, \omega_2, \alpha, \gamma)$
and $\Ex[D_{ij} | X_i, \tilde{X}_j, \phi] = \sigma(\langle \omega_1, X_i \rangle + \langle \omega_2, \tilde{X}_j \rangle + \alpha_i + \gamma_j)$,
where $\sigma(\cdot)$ takes a simple parametric form such as the sigmoid function. 
Some notable works in the MAR literature include \cite{Liang2016, WangNeurips2018, WangAAAI2018, wang2019}.

{\bf Missing not at random (MNAR).}
MNAR is the most challenging missingness model in matrix completion with a comparatively sparser literature.
In its fullest generality, in MNAR the following conditions are allowed:
(i) $\bD$ can depend on $\bY$ and other unobserved variables;
(ii) $D_{ij}$ can be correlated with $D_{ab}$ for all $(i,j) \neq (a,b)$;
(iii) $\min p_{ij} = 0$.
The first condition implies that $\bD$ and $\bY$ remain dependent even conditional on observed covariates.
The second condition allows the revelation of one outcome to alter the probability of another outcome being revealed. 
Finally, the third condition can restrict certain outcomes from ever being revealed. 
Hence, the literature has thus far only considered a {\em limited} version of MNAR with conditions cf. \cite{ma2019missing, bhattacharya2021matrix, yang2021tenips}.
In particular, they continue to make the following assumptions: $p_{ij}$ is a (nice) function solely of latent factors associated with entry $(i, j)$; each entry of $\bD$ is an independent (not necessarily identically distributed) Bernoulli r.v. with a strictly positive probability of being revealed, which are the assumptions as in MAR.
These assumption are what allow the weighted ERM framework of \eqref{eq:erm_weighted} to continue being valid. 
The methods proposed in \cite{ma2019missing, bhattacharya2021matrix, yang2021tenips} work for this {\em limited} MNAR setting by positing that $\bP$ is (approximately) low-rank, and recovers $\bP$ from $\bD$ via matrix completion algorithms. 
This is a generalization of the MAR setting as such an approach circumvents the requirement of meaningful auxiliary features $\bX$ to conduct propensity score estimation.
Additional works within the MNAR literature include 
\cite{zhu2019high, sportisse2020imputation_low_rank, sportisse2020estimation_PCA, causal_recommender_systems}.

As previously mentioned, our work operates under greater generality than the {\em limited} MNAR regime thus far considered in the literature. 
More specifically, our framework allows $\bD$ and $\bY$ to be dependent, provided $\bD \independent \bY | \bA$, and for $\bD$ to be any arbitrary function of $\bA$, subject to suitable observation patterns. 
We also allow for conditions (ii) and (iii) described above to hold, i.e., the entries in $\bD$ can be highly correlated and the minimum probability of observation can be deterministically set to $0$. 
In Section~\ref{sec:causal_framework}, we will formally introduce our causal framework to rigorously discuss these properties.

{\bf Summary of matrix completion results.}
Across the various models of missingness, the key theoretical results for low-rank matrix completion typically have error bounds that scale in the following form (see \cite{ieee_matrix_completion_overview}): 
\begin{align}
\frac{1}{mn}\| \bhA - \bA \|_F^2 = O\left(\frac{1}{\text{poly}(p_\min)} \cdot \frac{\text{poly}(r)}{\min(m,n)^{1-\delta}}\right)
\end{align}
for $\delta \geq 0$ and where $\text{poly}(\cdot)$ denotes polynomial dependence.
Here, $p_\min = \min p_{ij}$ and $r$ refers to the (approximate) rank of $\bA$. 
The most studied metric in the literature is the average error across all entries, $(1/mn) \| \bhA - \bA \|^2_F$,
though recent works have begun to analyze stronger metrics such as the maximum average error across all columns, $(1/m) \| \bhA - \bA \|^2_{2, \infty}$ (e.g., \cite{agarwal2019robustness, agarwal2020principal, agarwal2021causal}),
and the maximum entry-wise error, $\| \bhA - \bA \|_\max$ (e.g., \cite{LeeLiShahSong16}). 
Crucially, all of these error bounds scale with the inverse of $\text{poly}(p_\min)$. 
As discussed above, this immediately rules out settings where $p_\min = 0$, i.e., condition (iii) of MNAR above.
Finally, we remark that the literature studying the asymptotic properties of $\| \bhA - \bA \|$ (e.g., proving asymptotic normality) is relatively small.
Some notable works on the asymptotic analyses of matrix completion estimators under MCAR include \cite{chen2019inference, cai2020uncertainty, bhattacharya2021matrix}.

\subsection{Panel Data and Matrix Completion}\label{sec:panel_data_review}
In Section~\ref{sec:causal_framework}, we propose a causal framework for matrix completion that draws inspiration from
the rich and growing literature in econometrics on panel data and matrix completion; some relevant works include \cite{RSC, mRSC, arkhangelsky2019synthetic, bai2019matrix, fernandez2020low, athey2021matrix, agarwal2019robustness, agarwal2021synthetic, agarwal2021causal}.
As is common in matrix completion, these works impose a (approximate) low-rank factor model on the signal matrix (i.e., $\bA$), also known as an interactive fixed effects model, to capture structure across units and time (i.e., the rows and columns of the matrix, respectively).

{\bf Panel data \& matrix completion: an overview.}
As described in Section~\ref{sec:intro}, the sparsity structure considered in these works is one where for each row $i$, there is a column $j_i \in [n]$ such that $D_{i j} = 1$ for all $j < j_i$ and $D_{i j} = 0$ for $j \geq j_i$.
That is, all entries for a given row $i$ are observed till some column $j_i$, after which they are all missing.
The motivation for such a sparsity pattern comes from socio-economic policy making where $Y_{ij}$ represents unit $i$'s potential outcome at time step $j$ under ``control'', i.e., if no socio-economic intervention has yet been applied on unit $i$.
The time steps $[1, j_i -1]$ represent the period when unit $i$ is under control, and time steps $[j_i, n]$ represent the period when unit $i$ has undergone an intervention.
Hence, $Y_{ij}$ for $j > j_i$ is missing and the goal is to estimate the counterfactual of what would have happened to unit $i$ had it remained under control during $[j_i, n]$.
This particular setting is also known in the econometrics literature as ``synthetic contorls'' \cite{abadie1, abadie2}.
The statistical/causal parameter that is most commonly studied is for a ``treated'' unit $i$, to estimate
$
\frac{1}{n - j_i} \sum^n_{j = j_i} Y_{ij}.
$
That is, the average potential outcome of unit $i$ under control during the ``post-intervention'' period.
Most of these works make the additional assumption that each unit either remains under control for the entire time period under consideration, or undergoes an intervention at a time step that is common across all units.
\cite{athey2021matrix} is one notable work that allows for different post-intervention periods for each unit.

{\bf Connections to matrix completion with MNAR data.}
An attractive quality of this literature is that in some ways it allows for more relaxed conditions on $\bD$ and $\bP$ than those considered in the matrix completion with MNAR data literature discussed earlier, see~\cite{ma2019missing, yang2021tenips, sportisse2020estimation_PCA, sportisse2020imputation_low_rank, wang2019}.
In particular, the panel data literature allows the entries of $\bD$ to be correlated, e.g., if $D_{i j} = 0$, then $D_{i j'} = 0$ for $j' > j$.
Further, $\min p_{ij}$ is allowed to be $0$ and $j_i$ is allowed to depend on $\bA$.
On the other hand, the sparsity pattern considered in the panel data literature is far more restrictive compared to the works on matrix completion with MNAR data---as discussed above, in panel data settings, all columns for a given row are observed till a specific point, after which they are all missing (i.e.,  $D_{i j} = 1$ for all $j < j_i$ and $D_{i j} = 0$ for $j \geq j_i$).
Note that this also implies that $\bP$ is low-rank.
Such a sparsity pattern is unrealistic for many important applications for matrix completion, including recommendation systems and sequential decision-making.
Further, it is not straightforward to see how the target statistical/causal parameter 
$
\frac{1}{n - j_i} \sum^n_{j = j_i} Y_{ij}
$
is particularly meaningful outside the synthetic controls literature.
Hence, our aim with this work is to combine the best of both worlds, where we: (i) allow entries of $\bD$ to be correlated; (ii) allow $\min p_{ij} = 0$; (iii) make no parametric assumptions about $\bP$; (iv) allow $\bP$ to not be low-rank; (v) allow for general missingness patterns in the matrix that includes MCAR data as a special case.
Further the target parameter we aim to estimate (in expectation) is {\em each entry} $Y_{ij}$ for every $(i, j)$ pair.
Also, by formally bridging the panel data literature to more classical applications of matrix completion such as recommendation systems, we hope this spurs further investigation into the unexplored connections between these two fields.

{\bf Comparison with synthetic interventions.}
Our proposed framework framework builds upon the recent work of \cite{agarwal2021synthetic}, called synthetic interventions (\SI).
\SI~is a causal inference method to do tensor completion with MNAR data, where the dimensions of the order-3 tensor of interest are units, measurements, and interventions.
That is, an entry $Y_{ijd}$ of the tensor considered in \SI~refers to the potential outcome of the $i$-th unit, its $j$-th measurement, under the $d$-th intervention.
Their setup can be made a special case of ours by effectively {\em flattening} the tensor into a matrix, where the rows of the induced matrix still correspond to units, but a column is a double index for a measurement and an intervention, i.e., the $(i, j, d)$-th entry of the tensor corresponds to the $(i, (j, d))$-th entry of the induced matrix.
Given this simple reduction, we generalize the framework, algorithm, and theoretical results in \cite{agarwal2021synthetic} in the following ways. 
First, we formally extend the \SI~framework, to recover matrices under more general missingness patterns than that considered in \cite{agarwal2021synthetic}.
Doing so allows us to apply our framework to a wider variety of applications such as recommender systems, while the \SI~framework was introduced in the context of personalized policy evaluation and synthetic A/B testing.
Third, this work establishes point-wise finite-sample consistency and asymptotic normality of our proposed \SNN~algorithm, which was absent in \cite{agarwal2021synthetic} with respect to the \SI~algorithm. 
Indeed, in the context of the panel data literature, establishing point-wise asymptotic normality for each unit, (intervention, time)-tuple is of independent interest.
%

% section 3: framework
\section{A Causal Framework for Matrix Completion}\label{sec:causal_framework}
In this section, we develop a formal causal framework for matrix completion with MNAR data. 
In Section~\ref{sec:framework_po}, we show how to causally interpret matrix completion with MNAR data using the language of potential outcomes in Section~\ref{sec:framework_po}. 
We then state and justify our assumptions in Section~\ref{sec:framework_main_assumptions},
define our causal estimand in Section~\ref{sec:framework_causal_parameters}, 
and present our identification result in Section~\ref{sec:framework_identification}.
%
% Before we begin, we recall the definitions and notations set in Section~\ref{sec:problem_setup} of $\bA, \bD, \bP, \bY, \btY$. 

\subsection{Potential Outcomes} \label{sec:framework_po} 
We follow the potential outcomes framework of \cite{neyman, rubin}.
In particular, we let the r.v. $Y_{ij} \in \Rb$, as defined in Section~\ref{sec:problem_setup}, denote the potential outcome associated with each pair $(i,j)$ if it is revealed.
%
% In other words, $Y_{ij}$ is a philosophical quantity representing an outcome that can potentially be observed. 
%
For instance, in the case of recommender systems, $Y_{ij}$ can be interpreted as the rating user $i$ {\em would have} given to item $j$ had they rated it.
In the context of healthcare for example, $Y_{ij}$ could represent patient $i$'s health metric of interest (e.g. heart rate) {\em had they} been given treatment $j$. 
Finally, in the case of panel data setting, as discussed in Section~\ref{sec:panel_data_review}, $Y_{i, (a, t)}$ can denote the metric of interest for unit $i$ (e.g. revenue generated, socio-economic indicator), if they {\em would have} received the $a$-th socio-economic policy at time step $t$; here, $(a, t)$ represents the $j$-th column. 

If $D_{ij} = 1$, then by \eqref{eq:problem_setup} we see that we actually do observe the $(i,j)$-th potential outcome, i.e. $\tY_{ij} = Y_{ij}$.
That is, in the language of potential outcomes, we can interpret $\bD$ as the matrix of intervention assignments. 
Through this perspective, we remark that \eqref{eq:problem_setup} is an implicit assumption that is known in the causal inference literature as ``consistency'' or ``stable-unit-treatment-value assumption'' (SUTVA). 
As discussed earlier, the fact that $\bY \notindep \bD$ (e.g. a user's preference for a movie can determine whether they rate it) means that the potential outcomes are not independent of the intervention assignments.
This dependence is known in the causal inference literature as ``confounding''.
Lastly, as alluded to earlier, we generalize the standard potential outcomes framework in that a given unit can receive multiple interventions.
Traditionally, it is assumed that a unit receives exactly one intervention.
However, in applications like movie recommendation systems, a user can ``intervene'' and rate multiple movies.
Lastly, this framework also generalizes panel data settings, as we allow each unit to receive different interventions at different time steps; as discussed earlier, it is typically assumed that units are in control for a period of time, and then some subset of units receive one intervention for the remaining time steps.

\subsection{Assumptions} \label{sec:framework_main_assumptions}
Below, we state our causal assumptions and then provide their corresponding interpretations. 
\begin{assumption}[Low-rank factor model]\label{assump:LFM} %[Low-rank]
For every pair $(i,j)$, let 
\begin{align}
Y_{ij} = \langle u_i, v_j \rangle + \varepsilon_{ij},
\end{align}
where $u_i, v_j \in \Rb^r$ are latent vectors. 
Equivalently, we say $\bY = \bU \bV^T + \bE$, where $u_i$ refers to the $i$-th row of $\bU \in \Rb^{m \times r}$, and $v_j$ refers to the $j$-th row of $\bV \in \Rb^{n \times r}$.
\end{assumption} 
\begin{assumption}[Selection on latent factors] \label{assump:mean_ind} % [Conditional mean independence]
We have that for any intervention assignment $\bD$,  
\begin{align}
    \Ex[ \bE | \bU, \bV, \bD] = 0
\end{align}
%$\Ex[ \bE | \bU, \bV] = \Ex[ \bE | \bU, \bV, \bD] = 0$. 
% $\Ex[ \bE | \bU, \bV] = 0$ and $\bE \independent \bD ~|~ \{\bU, \bV\}$.
\end{assumption} 
{\em Neighbourhood rows and columns.}
For the remainder of this work, for a given column $j$, we refer to $\NR(j) =  \{a \in [m]: D_{a j} = 1\}$ as ``neighborhood rows'', i.e., rows where entries in column $j$ are not missing.
Similarly, for a given row $i$, we refer to $\NC(i) = \{b \in [n]: D_{ib} = 1\}$ as  ``neighborhood columns'', i.e., columns where entries in row $i$ are not missing. 
See Figure \ref{fig:s1} for a visual depiction of $\NR(j)$ and $\NC(i)$.
\begin{assumption}[Linear span inclusion] \label{assump:linear_span}  %[Collaborative filtering]
Conditioned on $\bD$, for a given pair $(i, j)$ and any $\Ic \subseteq \NR(j)$, if $|\Ic| \ge \mu$, then  $u_i$ lies in the linear row span of $\bU_{\Ic}$, i.e., there exists a $\beta \in \Rb^{| \Ic |}$ such that
\begin{align}
u_i = \sum_{\ell \in \Ic} \beta_\ell u_\ell
\end{align}
\end{assumption}
{\bf Interpretation of Assumptions~\ref{assump:LFM} to~\ref{assump:linear_span}} 
By the tower law, Assumption~\ref{assump:mean_ind} implies that $\Ex[ \bE | \bU, \bV] = 0$.
This together with Assumption~\ref{assump:LFM} posits that $\Ex[\bY | \bU, \bV]$ is a low-rank matrix with rank $r$.
As discussed in Section~\ref{sec:related_works},
this is a standard assumption within the matrix completion literature. 
Next, we remark that Assumption~\ref{assump:mean_ind}, coupled with Assumption~\ref{assump:LFM}, implies that 
\begin{align}
\Ex[\bY ~|~ \bU, \bV] = \Ex[\bY ~|~ \bU, \bV, \bD].
\end{align}
That is, the potential outcomes are mean independent of the intervention assignments, conditioned on the latent row and column factors. 
This has been termed as ``selection on latent factors'', see \cite{agarwal2021synthetic}.
Similar conditional independence conditions have been explored in \cite{athey2021matrix, kallus2018causal}.
Lastly, given Assumption~\ref{assump:LFM}, it follows that Assumption~\ref{assump:linear_span} is rather mild. 
To see this, suppose $\text{span}(\{u_\ell: \ell \in \Ic\}) = \Rb^r$, i.e., $\text{rank}(\bU_{\Ic, \cdot}) = r$.
Then, Assumption~\ref{assump:linear_span} immediately holds as $u_i \in \Rb^r$.   
More generally, if the rows of $\bU$ are randomly sampled sub-gaussian vectors, then $\text{span}(\{u_\ell: \ell \in \Ic\}) = \Rb^r$ for any set $\Ic$ holds w.h.p., provided $\mu \ge r$ is chosen to be sufficiently large; see  \cite{vershynin2018high} for details.

\subsection{Target Causal Estimand} \label{sec:framework_causal_parameters}
Define 
\begin{align} \label{eq:A}
	\bA \coloneqq \Ex[\bY | \bU, \bV].
\end{align}
Note that given Assumptions~\ref{assump:LFM} and~\ref{assump:mean_ind}, the definition of $\bA$ in~\eqref{eq:A} is consistent with the definition of $\bA$ used in Section~\ref{sec:problem_setup}.
We are now equipped to define our target causal estimand, which is $A_{ij}$; for the remainder, of the paper we focus on a particular pair $(i, j)$, without loss of generality. 
Note given Assumptions~\ref{assump:LFM} and~\ref{assump:mean_ind}, we can write 
\begin{align}
    A_{ij} \coloneqq \Ex[Y_{ij} | u_i, v_j].
\end{align} 
In words, $A_{ij}$ translates as the expected potential outcome for the $(i,j)$-th pair, conditioned on its row and column latent vectors $(u_i, v_j)$.
For instance, returning to recommender systems, $A_{ij}$ represents the expected rating user $i$ would supply for item $j$, conditioned on the latent features that characterize user $i$ and item $j$. 
In panel data settings, letting $j = (a, t)$, $A_{ij}$ represents the potential outcome of unit $i$ had it received the $a$-th intervention at time step $t$.

\subsection{Identification} \label{sec:framework_identification}
The following identification results  establishes that each entry of $\bA$ can be learned from observable quantities, i.e., from $\btY$. 
Practically speaking, this means that matrix completion with MNAR data for any pair $(i, j)$ is possible.
\begin{theorem}\label{thm:identification}
Let Assumptions~\ref{assump:LFM} to~\ref{assump:linear_span} hold.
For a given pair $(i, j)$ and $\Ic \subseteq \NR(j)$ with $| \Ic | \ge \mu$, suppose $\beta$ defined with respect to $\Ic$ as in Assumption~\ref{assump:linear_span}, is known. 
Then,
\begin{align}
   A_{ij} = \sum_{\ell \in \Ic} \beta_\ell \Ex[\tY_{\ell j} ~|~ \bU, \bV, \bD].
    % A_{ij} = \langle \Ex[\btY_{\Ic, j}], \ \beta \rangle ~\Big|~ \bU, \bV, \bD. 
\end{align}
\end{theorem} 
{\bf Interpretation.} 
Theorem~\ref{thm:identification} states that despite the missingness pattern being MNAR, if Assumptions \ref{assump:LFM} to \ref{assump:linear_span} hold, and given knowledge of the linear model parameter $\beta$, the causal estimand $A_{ij}$ can be expressed in terms of quantities that can be estimated from observed data, namely $\Ex[\btY_{\Ic, j}]$; this is known in the causal inference literature as ``identification''.
Note, $\Ic$ is deterministic given $\bD$.  
The key requirement of the missingness pattern $\bD$ is that $\Ic \subseteq \NR(j)$ is sufficiently large, which is parameterized by $\mu$, i.e., we require $\mu \gg r$ where $r$ is the rank of $\bA$.
That is, the number of rows for which column $j$ is observed is sufficiently large.
Thus, Theorem~\ref{thm:identification} suggests that the key quantity that enables the recovery of $A_{ij}$ is $\beta$. 
In Section~\ref{sec:estimator}, we provide an algorithm to estimate $\beta$, which in turn, allows us to estimate $A_{ij}$.

% section 3: algorithm
\section{\SNN: Matrix Completion with MNAR Data} \label{sec:estimator}
In this section, we introduce an algorithm, synthetic nearest neighbors (\SNN), for matrix completion with MNAR data. 
Towards this, we introduce helpful notation that will be used for the remainder of this work. 
Again, without loss of generality, we consider imputing the $(i,j)$-th entry of the matrix . 

\begin{figure} [!ht]
	\centering 
	\begin{subfigure}[b]{0.22\textwidth}
		\centering 
		\includegraphics[width=\linewidth]
		{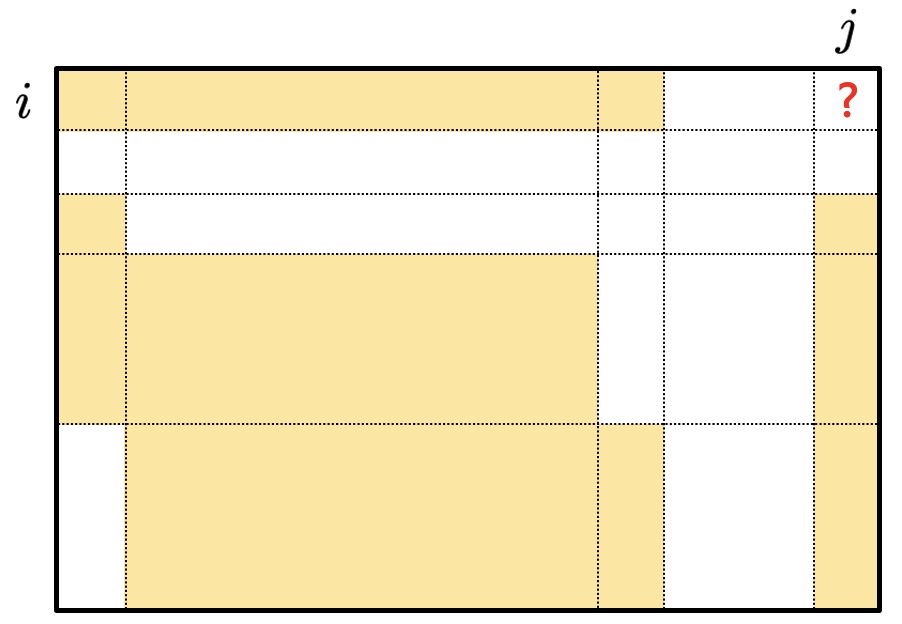}
		\caption{\smaller} 
		\label{fig:s0}
	\end{subfigure} 
	%\hfill 
	\begin{subfigure}[b]{0.26\textwidth}
		\centering 
		\includegraphics[width=\linewidth]
		{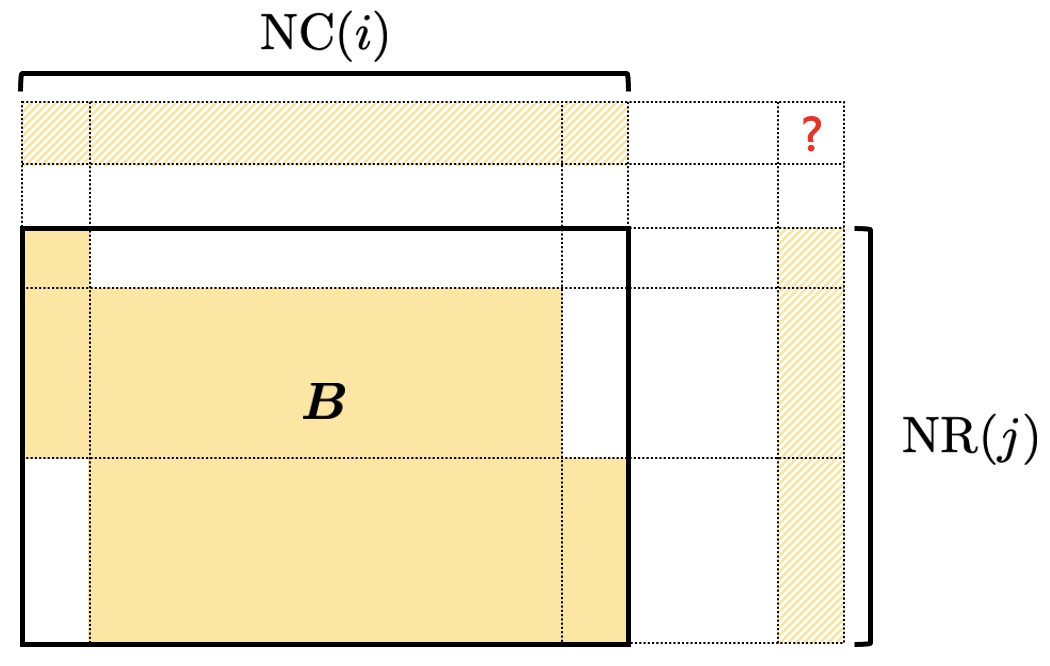}
		\caption{\smaller} 
		\label{fig:s1} 
	\end{subfigure} 
	%\hfill
		\begin{subfigure}[b]{0.24\textwidth}
		\centering 
		\includegraphics[width=\linewidth]
		{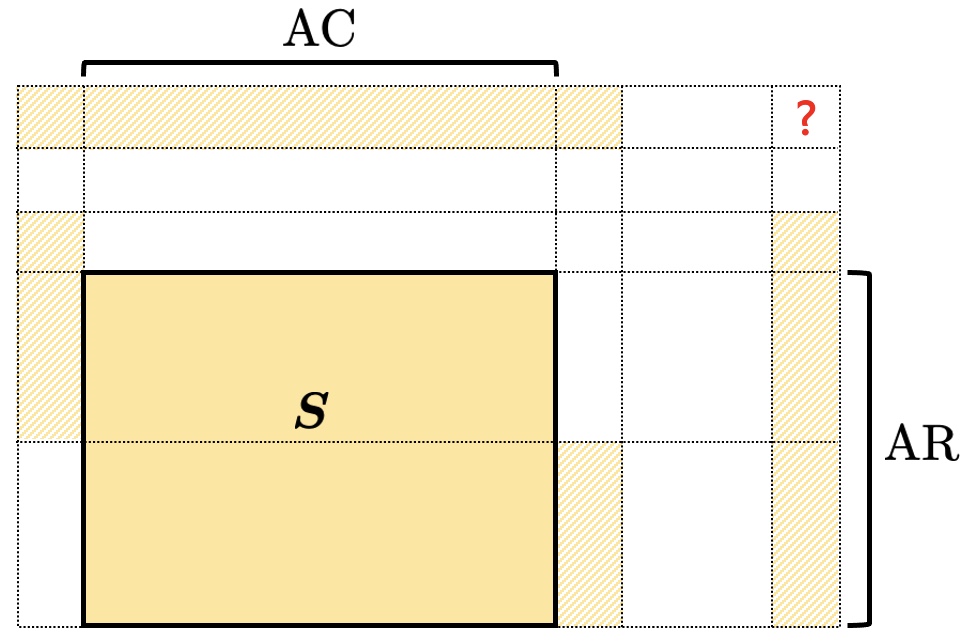}
		\caption{\smaller} 
		\label{fig:s2}
	\end{subfigure} 
	%\hfill 
	\begin{subfigure}[b]{0.26\textwidth}
		\centering 
		\includegraphics[width=\linewidth]
		{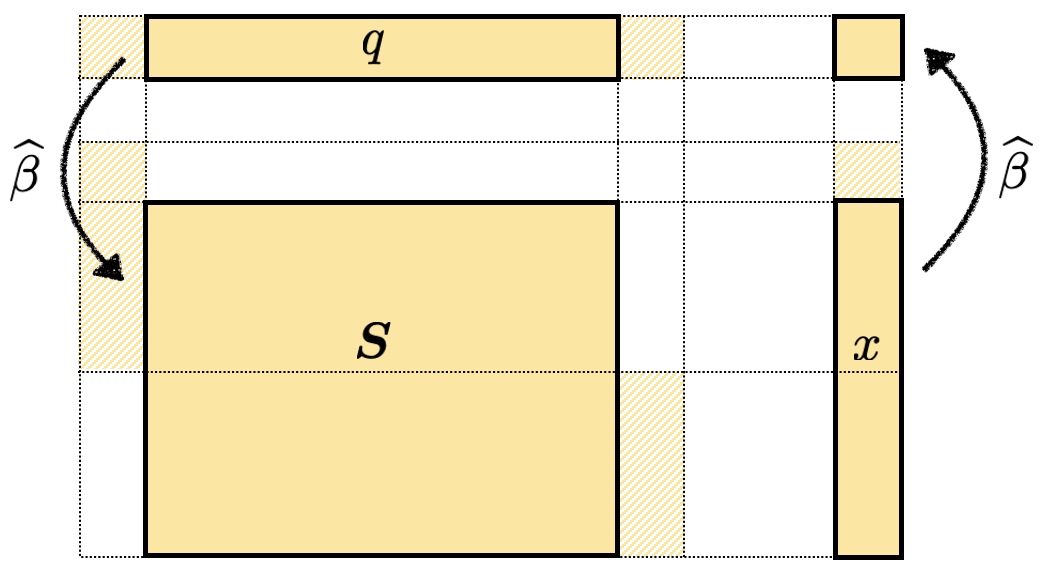}
		\caption{\smaller} 
		\label{fig:s3} 
	\end{subfigure} 
	\caption{We visually depict the various quantities needed to define the $\SNN$ algorithm.
	Figure \ref{fig:s0} depicts a particular sparsity pattern in our matrix $\btY$ with entry $(i, j)$ missing.
	Figure \ref{fig:s1} depicts $\NR(j)$ and $\NC(i)$.
	Figure \ref{fig:s2} depicts $\AR$, $\AC$, and $\bS$.
	Figure \ref{fig:s3} depicts the \SNN~algorithm with $K = 1$; for $K > 1$, we partition the rows in $S$ into $K$ mutually disjoint sets.
	}
	\label{fig:algo} 
\end{figure}

{\bf Notation.} 
Let $\AR \subseteq \NR(j)$ and $\AC \subseteq \NC(i)$ denote a subset of rows and columns, respectively, of $\btY$ that satisfy $D_{ab} = 1$ for all $(a,b) \in \AR \times \AC$. 
We refer to $\AR$ and $\AC$ as the ``anchor rows'' and ``anchor columns'' of pair $(i,j)$, respectively. 
Collectively, $\AR$ and $\AC$ form a fully observed sub-matrix of $\btY$; for ease of notation.
We refer to this $|\AR| \times |\AC|$ sub-matrix as $\bS \coloneqq [\tY_{ab}: (a,b) \in \AR \times \AC]$. 
See Figure \ref{fig:s2} for a visual depiction of $\AR$, $\AC$, and $\bS$.
Note, by construction $\bS$ is such that if entries from row $a$ are present in $\bS$, then $D_{a j} = 1$; similarly, if entries from column $b$ are present in $\bS$, then $D_{i b} = 1$
Additionally, let $q \coloneqq [\tY_{ib}: b \in \AC]$ and $x \coloneqq [\tY_{aj}: a \in \AR]$. 
$q \in \Rb^{|\AC|}$ refers to the columns in row $i$ which correspond to $\AC$; similarly, $x \in \Rb^{|\AR|}$ refers to the rows in column $j$ which correspond to $\AR$.
By construction, all the elements in $q$ and $x$ are not missing.
See Figure \ref{fig:s3} for a visual depiction of  $q$ and $x$.

\subsection{Algorithm} \label{sec:snn_alg}
We now present \SNN~in Algorithm~\ref{alg:snn} to impute the $(i,j)$-th entry. 
It has $K \in \Nb$ and $\lambda^{(k)} \in \Rb$ for $k \in [K]$ as hyper-parameters.

\begin{algorithm} [h!] 
\caption{\SNN$(i,j)$}
\label{alg:snn}
	\begin{algorithmic}
		\State Input: $\{\lambda^{(k)}: k \in [K]\}$,
		$\{(\AC^{(k)}, \AR^{(k)}): k \in [K]\}$ with mutually disjoint sets $\{\AR^{(k)}: k \in [K]\}$. 
%		 $\{\lambda_1, \dots, \lambda_K\}$, 
%		$\{(\AC^{(1)}, \AR^{(1)}), \dots, (\AC^{(K)}, \AR^{(K)})\}$ 
%		with mutually disjoint sets $\AR^{(1)}, \dots, \AR^{(K)}$
%		%$\AC$, mutually disjoint sets $\AR_1, \dots, \AR_K$
		\For {$k \in [K]$}
			\State 1. Define $\bS^{(k)} = [\tY_{ab}: (a,b) \in \AR^{(k)} \times \AC^{(k)}]$ 
			\State 2. Compute $\bS^{(k)} \leftarrow \sum_{\ell \ge 1} \htau^{(k)}_\ell  \hu^{(k)}_\ell \otimes \hv^{(k)}_\ell$
			\State 3. Compute 
			$\hbeta^{(k)} \leftarrow \Big( \sum_{\ell \le \lambda^{(k)}} (1/\htau_\ell^{(k)}) \hu^{(k)}_\ell \otimes \hv^{(k)}_\ell \Big) q^{(k)}$
			%$\hbeta^{(k)} \leftarrow \Big(\sum_{\ell \ge 1} \mathds{1}(\htau_\ell \ge \lambda^{(k)}) (1/\htau_\ell) ~ \hv_\ell \otimes \hu_\ell \Big) q$  
			\State 4. Compute $\hA_{ij}^{(k)} \leftarrow \langle x^{(k)}, \hbeta^{(k)} \rangle$ 
		\EndFor
		\State 4. Output $\hA_{ij} \leftarrow  \frac{1}{K} \sum_{k=1}^K \hA_{ij}^{(k)}$
	\end{algorithmic}
\end{algorithm} 
Note, for ease of notation, in Algorithm~\ref{alg:snn} we suppress the dependence on $i$ and $j$ in the definitions of $\{(\AC^{(k)}, \AR^{(k)}): k \in [K]\}$, $\bS^{(k)}$, $\hbeta^{(k)}$, $q^{(k)}$, and $x^{(k)}$.
That is, these quantities will change depending on which $(i,j)$-th entry of the matrix we aim to impute.
We continue to suppress this dependence for the remainder of the paper.
For a visual depiction of the \SNN~algorithm for $K = 1$, refer to Figure \ref{fig:s3}.
For $K > 1$, we simply re-run the \SNN~algorithm seperately for the $K$ disjoint subsets $\{\AR^{(k)}: k \in [K]\}$, and take the average of the estimates $\hA_{ij}^{(k)}$ for $k \in [K]$, produced by each iteration.

{\bf Interpretation.} 
\SNN~draws inspiration from the popular $K$ Nearest Neighbour (\knn) algorithm, described in Section \ref{sec:related_works}.
However, the key assumption underlying \knn~is that there {\em do exist} $K$ rows that are close to identical to the $i$-th row, with respect to some pre-defined metric.
However, it is not necessary that these $K$ rows exist {\em even} for a rank $1$ matrix.
As a simple example, consider a matrix $\bM \in \Rb^{m \times n}$ where $M_{i \cdot} = [i, 2i, \dots, ni]$.
% i.e., the $i$-th row of $\bM$ is simply a vector of $i$'s. 
%
By construction $\bM$ is rank $1$, but for any row, there does not exist any other row that is close to it in a mean squared sense; hence, it has no nearest neighbours.

The \SNN~algorithm overcomes this hurdle by first constructing $K$ ``synthetic'' neighbors of row $i$ from $\NR(j)$, where the $k$-th synthetic neighboring row is formed by a linear combination, defined by $\hbeta^{(k)}$, of the rows in $\AR^{(k)}$.
Then, similar to \knn, \SNN~estimates $A_{ij}$ by taking an average of the observed outcomes for column $j$ that are associated with the $K$ synthetic neighbors of row $i$. 
In words, $\hbeta^{(k)}$ is precisely the set of estimated linear weights that best recreates the observed outcomes of row $i$ from the rows in $\AR^{(k)}$, using observations from the columns in $\AC^{(k)}$.
This idea of matching rows via a linear re-weighting takes inspiration from the synthetic controls literature---see Section~\ref{sec:panel_data_review} for details.
To ensure the linear fit is appropriately regularized, a spectral sparsity constraint is imposed on $\bS^{(k)}$, which is parameterized by $\lambda^{(k)}$.
This constrained regression is known in the literature as principal component regression (PCR) (see~\cite{agarwal2019robustness, agarwal2020principal, agarwal2021synthetic, agarwal2021causal}). 
We note that in lieu of requiring that there exist $K$ close neighbouring rows as in \knn, \SNN~requires that the $i$-th row lies in the linear span of the rows in $\AR^{(k)}$; that is, given Assumption~\ref{assump:linear_span} holds, we require $|\AR^{(k)}| \ge \mu$.
Note that for the matrix $\bM$ described above, for any particular row, all other rows satisfy this linear span inclusion condition, i.e., $\mu = 1$ since $\bM$ is rank $1$. 

{\bf Choosing $\lambda^{(k)}$} 
There exist a number of principled heuristics to select the hyper-parameter $\lambda^{(k)}$, and we name a few here. 
As is standard within the statistics and ML literatures, the most popular data-driven approach is to use cross-validation. 
Another common approach is to use a universal thresholding scheme that preserves the singular values above a precomputed threshold (see \cite{donoho14, Chatterjee15}). 
Finally, a human-in-the-loop approach is to inspect the spectral characteristics of $\bS^{(k)}$ and choose $\lambda^{(k)}$ to be the natural ``elbow'' point that partitions the singular values into those of large and small magnitudes; 
in such a setting, the large magnitude singular values, which typically correspond to signal, are retained while the small magnitude singular values, which are often induced by noise, are filtered out. 
See the exposition on choosing the hyper-parameter for PCR in~\cite{agarwal2019robustness, agarwal2020principal, agarwal2021synthetic}.

{\bf Another Perspective on \SNN.}
\SNN~imputes $A_{ij}$ by building synthetic neighbors of row $i$ from $\NR(j)$. 
In Proposition~\ref{prop:snn_symmetry}, we demonstrate that $A_{ij}$ can be equivalently estimated by building synthetic neighbors of column $j$ from $\NC(i)$ through a simple ``transposition'' of Algorithm~\ref{alg:snn}. 

\begin{proposition} \label{prop:snn_symmetry} 
Consider any $k \in [K]$ and let $\hbeta^{(k)}$ be defined as in Algorithm~\ref{alg:snn}.
Further, let 
\begin{align}
\halpha^{(k)} =  (\sum_{\ell \le \lambda^{(k)}} (1/\htau^{(k)}_\ell) \hv^{(k)}_\ell \otimes \hu^{(k)}_\ell) x^{(k)}.
\end{align}
Then, 
\begin{align}
	\langle x^{(k)}, \hbeta^{(k)} \rangle = \langle q^{(k)}, \halpha^{(k)} \rangle. 
\end{align}
\end{proposition} 

\subsection{Finding Anchor Rows and Columns}  
Note the \SNN~algorithm takes as input $\{(\AC^{(k)}, \AR^{(k)}): k \in [K]\}$.
However, the question remains that given the matrix $\bD$, how to find these anchor rows and columns, with the additional constraint that  the $K$ set of anchor rows $\{\AR^{(k)}: k \in [K]\}$ are mutually joint.
In Section~\ref{sec:alg_anchor_algorithm}, we provide a practical algorithm $\anchor$ in Algorithm~\ref{alg:anchors} to find $\{(\AC^{(k)}, \AR^{(k)}): k \in [K]\}$ for a given pair $(i,j)$. 
In Section~\ref{sec:alg_anchor_applications}, we discuss some motivating applications where anchor rows and columns are naturally induced.

\subsubsection{Algorithmically Finding Anchor Rows and Columns via Maximum Biclique Search}\label{sec:alg_anchor_algorithm}
In particular, we reduce our task of finding anchor rows and columns to a well-known problem in the graph theory literature known as finding ``maximum bicliques''. 
We briefly explain how to do this simple reduction.
We first introduce some standard notation from graph theory.
Let $\Gc = (\Vc_1, \Vc_2, \Ec)$ denote a bipartite graph, where $(\Vc_1, \Vc_2)$ are the disjoint vertex sets and $\Ec \in \Vc_1 \times \Vc_2$ is the edge set, i.e., $(v_1, v_2) \in \Ec$ if there an edge between $v_1$ and $v_2$. 
Another way of representing $\Gc$ is via a bipartite incidence matrix $\bB \in \{0, 1\}^{|\Vc_1| \times |\Vc_2|}$ (or adjacency matrix).
In particular, $B_{ij} = 1$ if $(v_i, v_j) \in \Ec$.
If a sub-graph of $\Gc$ is complete, also called a biclique, then we denote it as $\BCc \subset \Gc$, i.e., there is an edge between any pair of nodes $(v_1, v_2) \in \BCc$. 
Now to see how to do the reduction between finding anchor rows and columns to the maximum biclique problem, recall $\bD \in \{0, 1\}^{m \times n}$ is our matrix of intervention assignments.
Note, $\bD$ immediately induces a bipartite graph with $|\Vc_1| = m$ and $|\Vc_2| = n$, i.e., the vertex sets $\Vc_1$ and  $\Vc_2$ correspond to the rows and columns of $\bD$, respectively.
We define $\Ec$ as follows, $(v_i, v_j) \in \Ec$ if $D_{ij} = 1$.
In other words, the incidence matrix $\bB \in \{0, 1\}^{m \times n}$ induced by this graph is exactly equal to $\bD$, i.e., $B_{ij} = 1$ if and only if $D_{ij} = 1$.

Given this reduction, we now describe how to practically implement the \anchor~algorithm. 
We assume access to two algorithms: \texttt{createGraph} and \texttt{maxBiclique}. 
The former, $\texttt{createGraph}: \bB \rightarrow \Gc$, takes as input a bipartite incidence matrix $\bB$ (or adjacency matrix) and returns a bipartite graph $\Gc$; 
we note that the Python package \texttt{NetworkX} is an excellent resource to generate such graphs. 
The latter, $\texttt{maxBiclique}: \Gc \rightarrow \{\BCc^{(\ell)}\}_{\ell \in [L]}$, takes as input a bipartite graph $\Gc$ and returns a set of $L$ maximal bicliques $\{\BCc^{(\ell)}\}_{\ell \in [L]}$; 
we refer the interested reader to \cite{biclique, biclique1, biclique2, biclique3} and references therein for example algorithms. 

\begin{algorithm} [!htb] 
\caption{\texttt{AnchorSubMatrix}$(i,j)$}
\label{alg:anchors}
	\begin{algorithmic}
		\State Input: \texttt{createGraph}, $\texttt{maxBiclique}$ 
		\State 1. Find $\NR(j)$ and $\NR(i)$ 
		\State 2. Assign $\bB \leftarrow [D_{ab} : (a, b) \in \NR(j) \times \NR(i)]$ 
		\State 3. Generate $\Gc \leftarrow \texttt{createGraph}(\bB)$
%		\State $\quad  $ (a) Assign $\bG \leftarrow
%		\left[ \begin{array} {cc} 
%			\boldsymbol{1} & \bB \\ 
%			\bB^T & \boldsymbol{1} 	
%			\end{array} 
%		\right]$
		\State 4. Compute $\{\BCc^{(\ell)} = (\Vc^{(\ell)}_1, \Vc^{(\ell)}_2, \Ec^{(\ell)})\}_{\ell \in [L]} \leftarrow \texttt{maxBiclique}(\Gc)$ 
		\State 5. Assign $\BCc^* = (\Vc^*_1, \Vc^*_2, \Ec^*) \leftarrow \argmax \min\{ |\Vc^{(\ell)}_1|, |\Vc^{(\ell)}_2|\}$ over $\ell \in [L]$ 
		\State 6. Output $\AR \leftarrow \Vc^*_1$ and $\AC \leftarrow \Vc^*_2$ 
		%\State 7. Output $\AC^{(1)}, \dots, \AC^{(K)} \leftarrow \AC$
		%\State 8. Output $\AR^{(k)} \leftarrow \AR$ for every $k$ 
	\end{algorithmic}
\end{algorithm}
Given $(\AC, \AR)$ from Algorithm~\ref{alg:anchors}, we can construct $\{(\AC^{(k)}, \AR^{(k)}): k \in [K]\}$ as follows: 
First, we assign $\AC^{(k)} \leftarrow \AC$ for every $k$, i.e., the anchor columns for each subgroup $k$ are all identically equal to $\AC$. 
Second, we (randomly) partition $\AR$ into $K$ subgroups of equal size and
then assign $\AR^{(k)}$ as the $k$-th subgroup of $\AR$ such that $|\AR^{(k)}| \sim |\AR|/K$; 
in doing so, we ensure that $\{\AR^{(k)}: k \in [K]\}$ are mutually disjoint sets. 
Note, for the purposes of theoretical analysis, we do not necessarily need to have $\AC^{(k)}$ be identical across all $K$. 
In Section~\ref{sec:theoretical_results}, we show how the estimation error of \SNN~scales with $|\AR^{(k)}|$ and $|\AC^{(k)}|$.
In short, our theoretical results suggest that we want $\{(\AC^{(k)}, \AR^{(k)}): k \in [K]\}$ to be large on average; 
It is sufficient that we choose $|\AC^{(k)}|, |\AR^{(k)}|$ such that $\min_{k \in [K]}\{|\AC^{(k)}|, |\AR^{(k)}| \}$ is as large as possible; this is essence what Step 5 of Algorithm \ref{alg:anchors} is doing.
%

% \begin{figure} [!ht]
% 	\centering 
% 	\begin{subfigure}[b]{0.4\textwidth}
% 		\centering 
% 		\includegraphics[width=\linewidth]
% 		{images/s0.jpg}
% 		\caption{\smaller .} 
% 		\label{fig:s0}
% 	\end{subfigure} 
% 	\hfill 
% 	\begin{subfigure}[b]{0.4\textwidth}
% 		\centering 
% 		\includegraphics[width=\linewidth]
% 		{images/s1.jpg}
% 		\caption{\smaller .} 
% 		\label{fig:s1} 
% 	\end{subfigure} 
% 	\\
% 		\begin{subfigure}[b]{0.4\textwidth}
% 		\centering 
% 		\includegraphics[width=\linewidth]
% 		{images/s2.jpg}
% 		\caption{\smaller .} 
% 		\label{fig:s2}
% 	\end{subfigure} 
% 	\hfill 
% 	\begin{subfigure}[b]{0.4\textwidth}
% 		\centering 
% 		\includegraphics[width=\linewidth]
% 		{images/s3.jpg}
% 		\caption{\smaller .} 
% 		\label{fig:s3} 
% 	\end{subfigure} 
% 	\caption{\smaller . 
% 	}
% 	\label{fig:algo} 
% \end{figure}

\subsubsection{Applications where Anchor Rows and Columns are Naturally Induced}\label{sec:alg_anchor_applications} %
In this section, we discuss the typical sparsity pattern in recommender systems and sequential decision-making paradigms, which include panel data settings, reinforcement learning, and sequential A/B testing.
We argue why these applications have a sparsity pattern where anchor rows and columns are naturally induced.

% \begin{figure}
% \centering
% \subcaptionbox{Coat \label{fig:coat}}
% %
% {\includegraphics[width=0.45\textwidth]{images/chen_coat.png}}
% %
% \hfill
% %
% \subcaptionbox{MovieLens-100k \label{fig:movie}}
% %
% {\includegraphics[width=0.45\textwidth]{images/chen_movie.png}}
% %
% \caption{}
% \end{figure}

\begin{figure}
	\centering 
	\begin{subfigure}[b]{0.3\textwidth}
		\centering 
		\includegraphics[width=\linewidth]
		{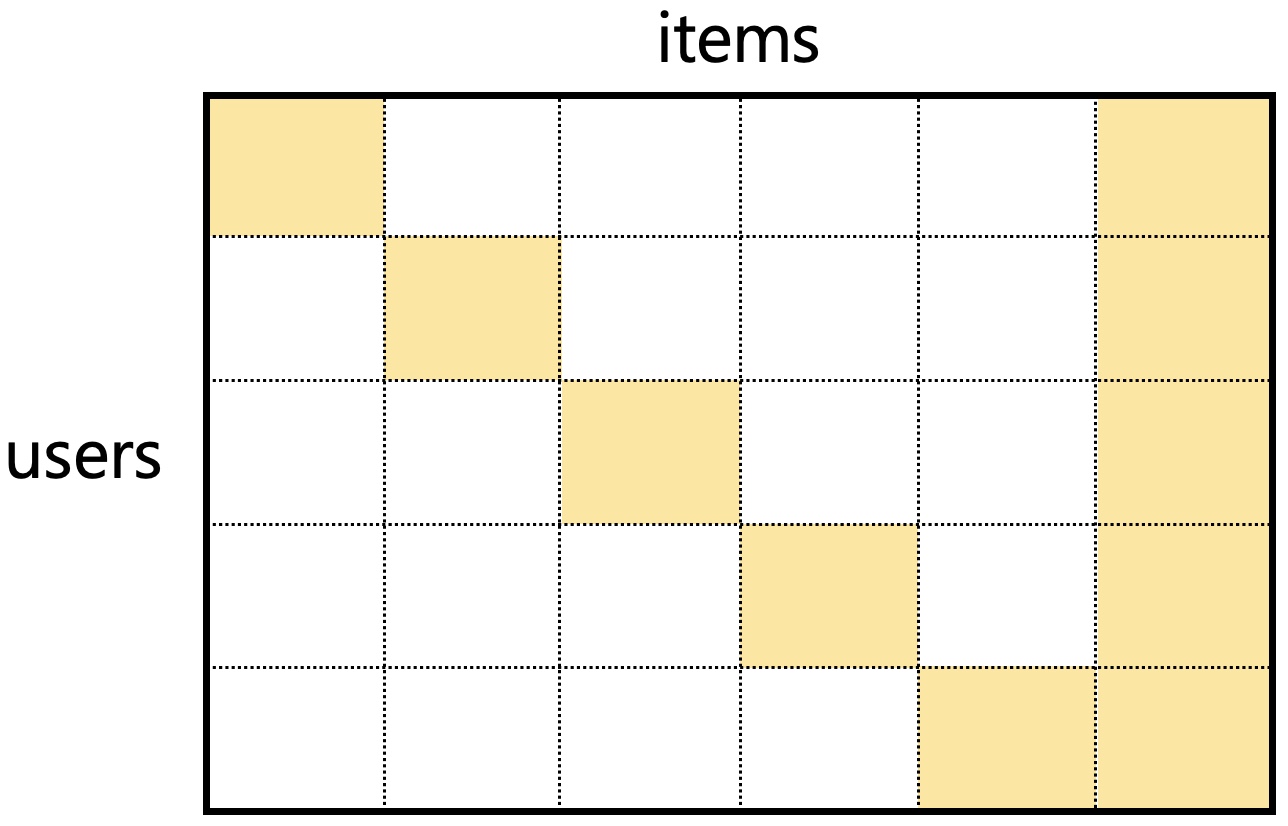}
		\caption{\smaller Recommender systems.} 
		\label{fig:staircase}
	\end{subfigure} 
	\hfill	\begin{subfigure}[b]{0.22\textwidth}
		\centering 
		\includegraphics[width=\linewidth]
		{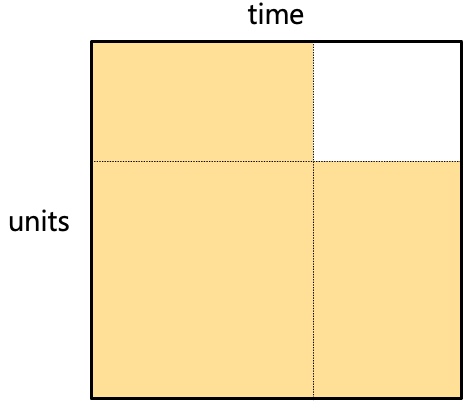}
		\caption{\smaller Panel data.} 
		\label{fig:panel} 
	\end{subfigure}
	\hfill 
	\begin{subfigure}[b]{0.33\textwidth}
		\centering 
		\includegraphics[width=\linewidth]
		{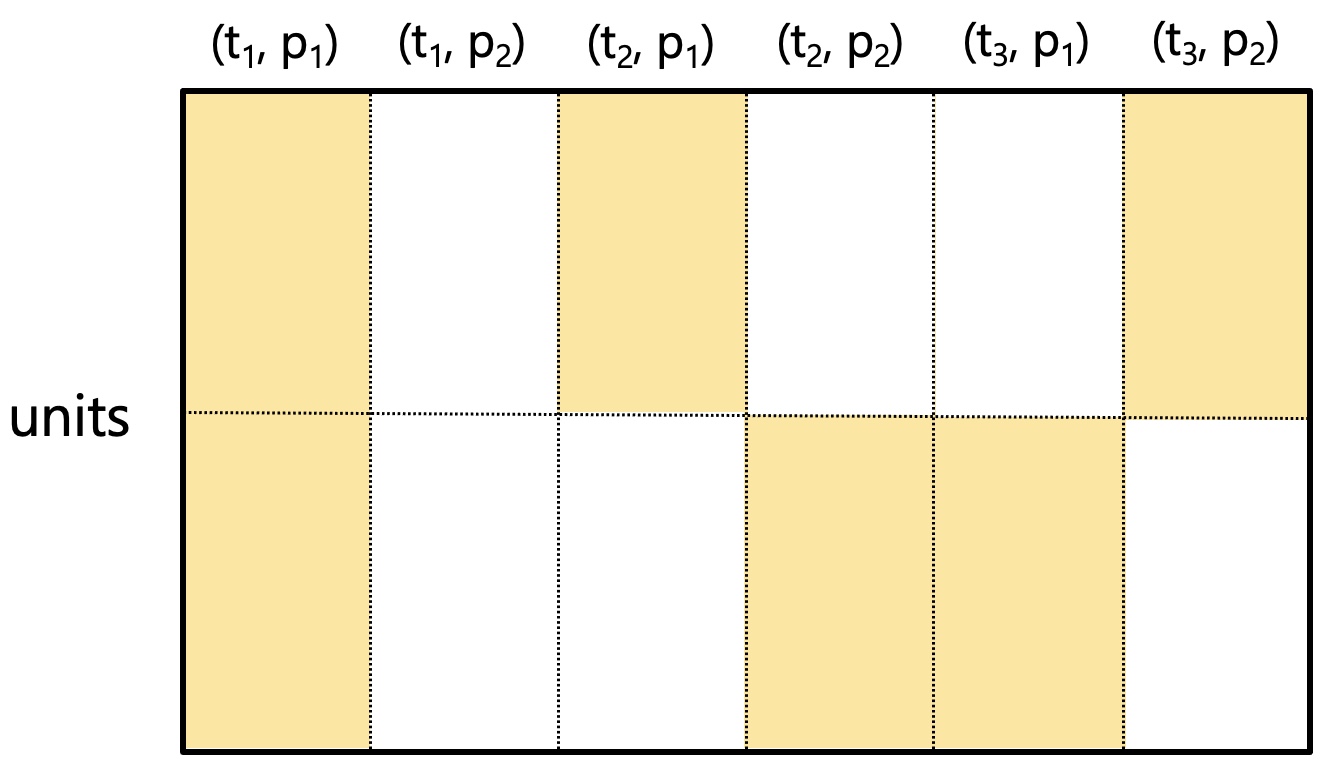}
		\caption{\smaller Sequential decision-making.} 
		\label{fig:sequential} 
	\end{subfigure} 
	\caption{\smaller In both \ref{fig:staircase}, \ref{fig:panel}, and \ref{fig:sequential}, observed entries are shown in yellow while unobserved entries are shown in white. Further, in \ref{fig:sequential}, the columns are indexed by (time, policy) tuples; here, $t_\ell$ and $p_\ell$ denote the $\ell$-th time period and policy, respectively. 
	}
	\label{fig:obs_patterns} 
\end{figure}

{\bf Recommender systems.} 
As stated earlier, one of the key motivating applications for matrix completion is recommender systems.
It has been noted in \cite{ma2019missing} that real-world recommender systems exhibit block-sparse structure; further the sparsity pattern is such that there is dependent missingness (i.e., $D_{ij} \notindep D_{ab}$ and zero probability of observing certain entries (i.e., $p_\min = 0$).  
% , which show the missingness matrices (i.e., $\bD$) for two well-studied datasets in the recommender system literature,  ``coat'' (\cite{schnabelfwang16}) and ``Movieles-100k'' (\cite{movielens}); here, black indicates an entry being revealed. 
% %
% In both Figure~\ref{fig:coat} and~\ref{fig:movie}, \cite{ma2019missing} show the missingness matrix on the left and identify the corresponding block structure using spectral biclustering on the right, which rearranges the rows and columns based on the biclustering result. 
% % 
% This sparsity pattern confirms the intuition that observations in recommender systems are often subject to selection-biases as users naturally consume items that they enjoy and avoid items that they dislike.
% %
% Indeed, from Figure~\ref{fig:coat} and~\ref{fig:movie} we see that $\bD, \bP$ are such that it is likely that $\min_{ij} p_{ij} = 0$ and that the entries of $\bD$ are dependent, thus violating the assumptions required to produce theoretical results for the current matrix completion with MNAR data literature (see Section~\ref{sec:related_works}).
% }
%
An extreme version of this selection-bias would induce a sparsity structure as shown in Figure~\ref{fig:staircase}. 
Within the context of movie recommender systems, a narrative for this missingness pattern is one where users only watch films that belong to genre(s) that they like and nothing else. 
However, in many recommender system applications, there exists a dense sub-matrix which corresponds to items that all users commonly rate---this corresponds to the rightmost columns of Figure~\ref{fig:staircase}.
This could occur if say a platform ask new users to indicate a subset of films that they enjoy. 
Indeed, this is a common practice for online platforms such as {\em Hulu, Netflix, StitchFix} to quickly learn a new user's preferences in order to provide a ``warm-start'' to their recommendation engine.
Alternatively, many a time there are a small subset of iconic films (e.g., {\em Titanic} or {\em Star Wars}) that a large majority of users have watched.
In this example in Figure~\ref{fig:staircase}, all users can be used as anchor rows, and the set of items that are commonly rated across all users can be used as anchor columns. 
Further, we remark that in this example $\bP$ is not low-rank, thus violating the key assumption required to learn $p_{ij}$ in \cite{ma2019missing, cai2020uncertainty, bhattacharya2021matrix}.

% \denniscomment{$p_{ij}$ not $p_{ij}$}

%\anishcomment{transpose Figure~\ref{fig:staircase} to match description above.}

{\bf Sequential decision-making.} 
As described earlier, in sequential decision-making, data is collected across units (e.g., individuals, customer types, geographic locations) over time in a sequential manner, where each unit is likely to be observed under a single or small set of interventions out of many at any time period. 
Many sequential decision-making problems can be phrased this way, including
(i) panel data settings in econometrics;
(ii) reinforcement learning and its variants (e.g., online learning, contextual bandits); 
(iii) sequential A/B testing.
In (ii), an intervention denotes both the action picked and the observed state for that given time period; 
meanwhile in (iii), platforms run experiments on different customer types in a sequential and/or adaptive manner over time.
The induced matrix in these settings has rows index units and columns index time-intervention pairs.
%
% %
% An important goal in such scenarios is to impute this induced matrix, i.e., estimate the counterfactual outcome of what would have happened to each unit, at each time period, under all interventions of interest.
%
% We now discuss a setting in which anchor rows and columns are naturally induced in such applications.
%
It is common in many of these sequential decision-making settings that there is a time period when all units are under the same intervention.
This is usually done to collect ``control'' data about each unit to establish its baseline.
For example, in an e-commerce setting, companies commonly estimate the baseline engagement level of a customer to understand the treatment effect of a discount policy; 
similarly, in clinical trials, pharmaceutical companies collect health metrics of patients to establish the treatment effect a particular therapy has.
Further, the assumption that such a control period exists is standard in the synthetic controls literature.
For an illustration of the sparsity pattern in the induced matrix with a control period, see Figure~\ref{fig:panel}.
Hence, this ``control'' period in sequential decision-making can serve as our anchor columns, and all units can serve as anchor rows.

% section 4: theoretical results
\section{Theoretical Results} \label{sec:theoretical_results} 
Below, we establish the statistical properties of the \SNN~algorithm. 
Without loss of generality, we consider a specific pair $(i,j)$.
Recall from our discussion earlier, we suppress dependencies on $(i,j)$, e.g., all anchor rows and columns $\AR^{(k)}, \AC^{(k)}$ are defined with respect to $(i,j)$.
In Section~\ref{sec:additional_assumptions}, we state additional assumptions required to establish the theoretical results.
In Sections~\ref{sec:finite_sample_consistency} and~\ref{sec:asymptotic_normality}, we establish finite-sample consistency and asymptotic normality of the \SNN~algorithm for a given entry $(i, j)$.
In Sections~\ref{sec:discussion_assumption} and~\ref{sec:discussion_results}, we discuss our assumptions and theoretical results, respectively.

{\bf Notation.} 
For every vector $v \in \Rb^a$, let  $\| v \|_p$ denotes its $\ell_p$-norm. 
For the remainder of this work, let $\Ec = \{\bU, \bV, \bD\}$, i.e., the collection of latent factors and the observed missingness pattern.
Recall the definition of $\bS^{(k)}$, $\AR^{(k)}$ and $\AC^{(k)}$ from Section \ref{sec:snn_alg}.
Moreover, for every $k \in [K]$, we denote the SVD of $\Ex[\bS^{(k)} ~|~ \Ec]$ as  
\begin{align}
	\Ex[\bS^{(k)} ~|~ \Ec] = \sum_{\ell=1}^{r^{(k)}} \tau^{(k)}_\ell u^{(k)}_\ell \otimes v^{(k)}_\ell;  
\end{align}
here, $r^{(k)} = \text{rank}(\Ex[\bS^{(k)} ~|~ \Ec])$. 
We denote $\bU^{(k)} \in \Rb^{|\AR^{(k)}| \times r^{(k)}}$ and $\bV^{(k)} \in \Rb^{|\AC^{(k)}| \times r^{(k)}}$ as the matrices of left and right singular vectors, respectively, i.e., $u^{(k)}_\ell \in \Rb^{|\AR^{(k)}|}$ and $v^{(k)}_\ell \in \Rb^{|\AC^{(k)}|}$ form the $\ell$-th columns of $\bU^{(k)}$ and $\bV^{(k)}$, respectively. 

\subsection{Additional Assumptions}\label{sec:additional_assumptions}
We state additional assumptions required to establish guarantees for the \SNN~algorithm.
In Section~\ref{sec:discussion_assumption} we provide interpretations for Assumptions~\ref{assump:spectra} and \ref{assump:subspace};
Assumptions~\ref{assump:subg} and \ref{assump:bounded} are relatively standard and self-explanatory.
Below, $k$ is indexed over $[K]$, where recall $K$ is a hyper-parameter of the \SNN~algorithm.

\begin{assumption}[Sub-gaussian noise] \label{assump:subg}
Conditioned on $\Ec$, $\varepsilon_{ij}$ are independent sub-gaussian mean-zero r.v.s with $\Ex[\varepsilon_{ij}^2] = \sigma_{ij}^2 \le \sigma^2$ and $\| \varepsilon_{ij} \|_{\psi_2} \le C \sigma_{ij}$ for some constants $C > 0$ and $\sigma > 0$. 
\end{assumption} 

\begin{assumption}[Bounded expected potential outcomes] \label{assump:bounded}
Conditioned on $\Ec$, $A_{ij} \in [-1,1]$.\footnote{The precise bound $[-1,1]$ is without loss of generality, i.e., it can be extended to $[a,b]$ for any $a, b \in \Rb$ with $a \le b$.} 
\end{assumption} 

\begin{assumption}[Well-balanced spectra]\label{assump:spectra}
Conditioned on $\Ec$ and given a pair $(i,j)$ as well as subgroup $k$, the $r^{(k)}$ nonzero singular values $\tau_\ell^{(k)}$ of $\Ex[\bS^{(k)} ~|~ \Ec]$ are well-balanced, i.e., there exist universal constants $c, c' > 0$ that satisfy
$$\tau^{(k)}_{r^{(k)}} / \tau^{(k)}_1 \ge c, \quad \| \Ex[\bS^{(k)} ~|~ \Ec] \|_F^2 ~\ge c' |\eAC^{(k)}| \cdot | \eAR^{(k)} |.$$ 
\end{assumption} 

\begin{assumption}[Subspace inclusion]\label{assump:subspace}
Conditioned on $\Ec$ and given a pair $(i,j)$ as well as subgroup $k$,
$$\Ex[x^{(k)} ~|~ \Ec] \in \emph{colspan}(\Ex[\bS^{(k)} ~|~ \Ec]),$$
where we recall $x^{(k)}$ is defined in Section \ref{sec:snn_alg}.
\end{assumption} 

% consistency
\subsection{Finite-sample Consistency}\label{sec:finite_sample_consistency}
The following result establishes that the \SNN~algorithm outputs entry-wise consistent estimates of $\bA$, i.e., we establish consistency in $\| \cdot \|_\max$-norm. 
To simplify notation, we will henceforth absorb dependencies on $\sigma$ into the constant within $O_p(\cdot)$.
That is, we assume there exists an absolute constant $C \ge 0$ such that $\sigma \le C$.
\begin{theorem} \label{thm:consistency}
Conditioned on $\Ec$, for a given pair $(i,j)$ and subgroup $k \in [K]$, suppose $|\eAR^{(k)}| \ge \mu$ and let Assumptions~\ref{assump:LFM} to \ref{assump:subspace} hold.
Further, let $K = o( \min_k |\eAC^{(k)}|^{10} |\eAR^{(k)}|^{10} )$. 
Finally, for each $k$, let $\lambda^{(k)} = \text{rank}(\Ex[\bS^{(k)}])$, where $\lambda^{(k)}$ is defined as in Algorithm~\ref{alg:snn}. 
Then,
\begin{align}
	\hA_{ij} - A_{ij} 
	&= O_p
	\left( 
	\frac{1}{K}\left\{
	\sum_{k=1}^K \frac{(r^{(k)})^{1/2}}{ |\eAC^{(k)}|^{1/4}} 
	+  \sum_{k=1}^K \frac{(r^{(k)})^{3/2} \| \tbeta^{(k)}\|_1 \log^{1/2}(|\eAC^{(k)}| |\eAR^{(k)}|)} { \min\{ |\eAC^{(k)}|^{1/2}, |\eAR^{(k)}|^{1/2} \}}
	+ \left[ \sum_{k=1}^K \| \tbeta^{(k)} \|_2^2 \right]^{1/2} 
\right\} \right).
\end{align} 
where $\tbeta^{(k)} = \Pc_{U^{(k)}} \beta^{(k)}$ is the projection of $\beta^{(k)}$ onto the subspace spanned by the columns of $\bU^{(k)}$. We assume $\|\tbeta^{(k)}\|_2 \ge c$, for some absolute constant $c \ge 0$.
\end{theorem}

\begin{cor}\label{cor:consistency}
Suppose $|\AC^{(k)}|, |\AR^{(k)}| = N$ for all $k \in [K]$.
Let $\beta_{\max, 2} = \max_k \| \tbeta^{(k)} \|_2$,  $\beta_{\max, 1} = \max_k \| \tbeta^{(k)} \|_1$, and $r_\max = \max_k r^{(k)}$.
Let the setup of Theorem~\ref{thm:consistency} hold.
Then,
\begin{align}
\hA_{ij} - A_{ij} =
O_p\left( 
\frac{r_\max^{1/2}}{N^{1/4}}
+ \frac{r_\max^{3/2} \cdot \beta_{\max, 1} \cdot  \log^{1/2}(N)}{N^{1/2}}
+ \frac{\beta_{\max, 2}}{\sqrt{K}}
\right)
\end{align}
\end{cor}

Note, Theorem~\ref{thm:consistency} does not require $N \rightarrow \infty$ to establish consistency of the \SNN~estimator.
Rather, that $|\AR^{(k)}|, |\AC^{(k)}|$ is growing {\em on average} (ignoring logarithmic factors and dependence on $\beta^{(k)}, r^{(k)}, \sigma$).
However, we state Corollary~\ref{cor:consistency} to help further interpret our results in Section~\ref{sec:discussion_results}.

{\bf Implication for matrix completion with MCAR data.}
Proposition \ref{prop:MCAR} below shows that \SNN~provides uniform entry-wise consistency for matrix completion with MCAR data as a special case if $p$, the probability of observing an entry, is sufficiently large.
\begin{proposition}[\SNN~for matrix completion with MCAR data]\label{prop:MCAR}
Let the setup of Theorem \ref{thm:consistency} hold.
Further, let $m = n = L$.
Assume each entry $(i, j)$ is revealed with uniform probability $p \in (0, 1]$, independent of everything else.
Fix any $\delta > 0$.
Let  
$$
p \ge \left(\frac{ Q}{L}\right)^{\frac{1}{Q^2}}
$$
with $Q = C^* \delta^{-6}$, where $C^*$ is a function only of $\beta^{(k)}, r^{(k)}$ for $k \in [K]$, $\sigma$, and $\log(L)$.

Then with probability at least $1 - \frac{C}{L^8}$, where $C > 0$ is an absolute constant, there exists sufficient anchor rows and columns, $\AR^{(k)}, \AC^{(k)}$, such that uniformly for all $(i, j) \in [m] \times [n]$,
$$
\hA_{ij} - A_{ij} = O_p(\delta).
$$
Hence, for any fixed $p > 0$, we have that $\hA_{ij} - A_{ij} = o(1)$ uniformly for all $(i, j) \in [m] \times [n]$ as $L \to \infty$.
\end{proposition}

% normality
\subsection{Asymptotic Normality}\label{sec:asymptotic_normality} 
The following establishes that the entry-wise estimate $\hA_{ij}$ of the \SNN~algorithm is asymptotically normal around the target causal parameter $A_{ij}$. 
\begin{theorem} \label{thm:normality}
For a given pair $(i,j)$ and subgroup $k$, let the setup of Theorem~\ref{thm:consistency} hold. 
Define 
\begin{align}
(\tilde{\sigma}^{(k)})^2 := \sum_{\ell \in \eAR^{(k)}} (\tbeta^{(k)}_\ell \sigma_{\ell j})^2
\end{align}
Further, let the following conditions holds
\begin{itemize}
    \item[(i)] $K \rightarrow \infty$; 
    \item[(ii)] $|\eAC^{(k)}|, |\eAR^{(k)}| \rightarrow \infty$ for each $k$;
    \item[(iii)] $r^{(k)} \| \tbeta^{(k)} \|^2_1 \log(|\eAC^{(k)}| |\eAR^{(k)}|) = o(\min\{|\eAC^{(k)}|, |\eAR^{(k)}|\})$ for each $k$;
    \item[(iv)] 
    \begin{align}\label{eq:normality_cond1}
    \sum_{k=1}^K \left( \frac{(r^{(k)})^{1/2}}{ |\eAC^{(k)}|^{1/4}} + \frac{(r^{(k)})^{3/2}\| \tbeta^{(k)}\|_1 \log^{1/2}(|\eAC^{(k)}| |\eAR^{(k)}|)} { \min\{ |\eAC^{(k)}|^{1/2}, |\eAR^{(k)}|^{1/2} \}}\right)  
    = o\left( \left[\sum^K_{k = 1} (\tilde{\sigma}^{(k)})^2 \right]^{1/2}\right)
    \end{align}
\end{itemize}
% %
% (i) 
% %
% (ii)   
% %
% (iii) 
% %
% (iv) 
Then conditioned on $\Ec$, 
\begin{align}
	\frac{K(\hA_{ij} - A_{ij})}{  \left[\sum^K_{k = 1} (\tilde{\sigma}^{(k)})^2 \right]^{1/2}} \xrightarrow{d} 
	\mathcal{N} \left(0,  1  \right). 
\end{align}
\end{theorem}

\begin{remark}
Recall the notation in Corollary \ref{cor:consistency}.
Then one can easily verify a sufficient property for condition (iii) in Theorem \ref{thm:normality} is
\begin{align}
    r_\max \cdot \beta^2_{\max, 1} \cdot \log(N) = o(N)
\end{align}
Further, let $\tilde{\sigma}_\min = \min_k \tilde{\sigma}^{(k)}$.
Then one can easily verify a sufficient property for condition (iv) in Theorem \ref{thm:normality} is
\begin{align}
    K  
    = o\left(\tilde{\sigma}_\min \cdot
    \min \left\{ \frac{N^{1/2}}{r_\max}, \  \frac{N}{r_\max^{3} \cdot \beta^2_{\max, 1} \cdot \log(N)}\right\} \label{eq:normality_sufficiency}
    \right) 
\end{align}
If we ignore dependence on logarithmic factors and on $\beta_{\max, 1}, r_\max, \tilde{\sigma}_\min$, \eqref{eq:normality_sufficiency} essentially requires that 
\begin{align}
K = o(N^{1/2}).
\end{align}
Practically, this can be interpreted as saying that to ensure valid confidence intervals, the number of synthetic nearest neighbours, i.e., $K$, we construct in \emph{\SNN}~cannot scale too quickly relative to the number of anchor rows and columns, i.e., $|\eAR^{(k)}|, |\eAC^{(k)}|$.
\end{remark}

\subsection{Discussion of Assumptions}\label{sec:discussion_assumption}

{\bf Interpretation of Assumption \ref{assump:spectra}.}
Assumption~\ref{assump:spectra} requires that the nonzero singular values of $\Ex[\bS^{(k)} ~|~ \Ec]$ are well-balanced. 
Such an assumption is quite standard with the econometrics factor model and matrix completion literature.
For example, it is analogous to incoherence-style conditions; see Assumption A of \cite{bai2019matrix} and the discussion of theoretical results in \cite{agarwal2019robustness}. 
It is also closely related to the notion of pervasiveness, see Proposition 3.2 of \cite{fan2018eigenvector}.
Indeed, the assumption that there is a gap between the top few singular values of a matrix of interest, and the remaining singular values has been widely adopted in the econometrics literature of  large dimensional factor analysis dating back to \cite{chamberlainfactor}.
Crucially though, these works within econometrics (e.g. \cite{bai2019matrix}, \cite{fan2018eigenvector}, \cite{chamberlainfactor}) aim to accurately estimate the factors themselves, which require making additional assumptions about the spectra of the matrix of interest to ensure these factors are uniquely identifiable.
Instead we simply require that these low-rank factors exist, but do not explicitly require accurately estimating them.
Assumption \ref{assump:spectra} has also been shown to hold with high-probability for the canonical probabilistic generating process used to analyze probabilistic principal component analysis in \cite{bayesianpca} and \cite{probpca};
here, the observations are assumed to be a high-dimensional embedding of a low-rank matrix with independent sub-Gaussian entries (see Proposition 4.2 of \cite{agarwal2019robustness}). 
Within the matrix/tensor completion literature, for an overview of where the well-balanced spectra assumption is utilized, see \cite{cai2021nonconvex} and references therein.
Practically speaking, Assumption~\ref{assump:spectra} can be empirically validated by plotting the spectrum of $\bS^{(k)}$, defined in Algorithm~\ref{alg:snn}; if there is a natural ``elbow'' point in the singular spectrum of $\bS^{(k)}$, i.e., there are a relatively small number of singular values that have a large and approximately equal magnitude, and the remaining singular values are significantly smaller, then Assumption~\ref{assump:spectra} is likely to hold.
For further discussion of this empirical robustness check, please refer to the related discussion in \cite{agarwal2021synthetic}. 

{\bf Interpretation of Assumption \ref{assump:subspace}.}
Recall from Algorithm~\ref{alg:snn} that we learn the model $\hbeta^{(k)}$ by regressing $q^{(k)}$ on $\bS^{(k)}$.
$\hA_{ij}^{(k)}$ is then estimated by applying the model $\hbeta^{(k)}$ on the outcomes in $x^{(k)}$ (i.e., the entries in the $j$-th column of the rows $\AR^{(k)}$).
The key question that remains is why would a model learned between $q^{(k)}$ and $\bS^{(k)}$, {\em generalize} well to accurately estimate $A_{ij}^{(k)}$ using $\langle x^{(k)}, \hbeta^{(k)} \rangle$.
Normally, in statistical learning, such generalization requires making distributional assumptions about the training data (i.e.,  $\bS^{(k)}$) and the testing data (i.e., $x^{(k)}$).
For example, each column of $\bS^{(k)}$ and $x^{(k)}$ are sampled i.i.d.
However, we do not want to make such an assumption as it is unrealistic in setting such as recommendation systems, e.g., the ratings users give different movies is likely to be neither identically nor independently distributed.
Indeed, by conditioning on $\Ec$, we are implicitly conditioning on $\bU$ and $\bV$, which requires our analysis to be {\em instance dependent}, i.e., has to hold for the specific sampling of the latent factors $\bU$ and $\bV$.
To circumvent making any distribution assumptions, we make the natural assumption that in expectation, $x^{(k)}$ lies within the linear span of $\bS^{(k)}$.
Such a condition is necessary as well for generalization, e.g., if every entry of $\Ex[\bS^{(k)} ~|~ \Ec]$ is equal to $0$, then no meaningful model $\hbeta^{(k)}$ can learned.
Such an assumption has also been explored in \cite{agarwal2020principal, agarwal2021synthetic, agarwal2021causal}.
In particular, in \cite{agarwal2021synthetic} the authors provide a data-driven hypothesis test to verify when such a condition holds.

\subsection{Discussion of Results}\label{sec:discussion_results}
To ease the discussion of the interpretation of the results, we will ignore dependence on logarithmic factors, and $\beta^{(k)}, r^{(k)}, \sigma$.

{\bf Sample complexity.}
Note that even if $D_{ij} = 1$, estimating $A_{ij}$ is not straightforward;
we never get to observe $A_{ij}$, rather we only observe $Y_{ij}$, where $Y_{ij} = A_{ij} + \varepsilon_{ij}$.
That is, even if $D_{ij} = 1$, our observation of $A_{ij}$ is corrupted by noise and we only get a single sample of it.
Remarkably, despite having access to (at most) a single noisy sample of $A_{ij}$, the estimate $\hA_{ij}$ produced by the \SNN~algorithm is consistent and asymptotically normal around $A_{ij}$.
Of course, this is assuming a low-rank factor model and a suitable observation pattern.
Hypothetically, if we get $K$ independent noisy samples of $A_{ij}$, denoted by $Y^{(1)}_{ij}, \dots, Y^{(K)}_{ij}$, the maximum likelihood estimator would be the empirical mean, $\frac{1}{K} \sum^K_{k = 1} Y^{(k)}_{ij}$.
In this hypothetical scenario, this empirical mean would concentrate around $A_{ij}$ with error scaling as $O(K^{-1/2})$, i.e., with this estimation procedure to obtain an additive error of $O(\delta)$, we would need $K = \Omega(\delta^{-2})$ independent copies.

Now in comparison to the hypothetical scenario above where we have access to $K$ independent samples, Corollary~\ref{cor:consistency} effectively establishes that with access to at most $N^2 \times K$ observations, the error of the \SNN~estimator scales as $O(\max(N^{-1/4}, K^{-1/2}))$; this is assuming $|\AC^{(k)}|, |\AR^{(k)}| = N$ for all $k \in [K]$, as in Corollary \ref{cor:consistency}.
This implies for any $\delta > 0$, $A_{ij}$ can be estimated to within an additive error of $O(\delta)$, if  $N = \Omega(\delta^{-1/4})$ and $K = \Omega(\delta^{-2})$.
Hence, compared to if we had $K$ independent noisy samples of each $A_{ij}$, we pay an additional cost of $N^2$ in terms of the number of samples needed, and $N^{-1/4}$ in terms of the estimation error rate even though we either do not observe a sample of $A_{ij}$ (i.e., it is missing), or only observe a single, noisy instantiation of it in $\tY_{ij}$. 
% only having access to a single, possibly missing, sample of $A_{ij}$ (denoted by $\tY_{ij}$). 
That is, to obtain estimation error of $O(\delta)$, it requires $O(\delta^{-2} \times \delta^{-4})$ observations (across different entries). 
%
% Again, we note that our bounds do not actually depend on $N$ and $N$; rather they depend on $|\AR^{(k)}|,|\AC^{(k)}|$ on average over $k \in [K]$. 

\begin{figure}[ht!]
	\centering 
	\includegraphics[width=0.3\linewidth]
	{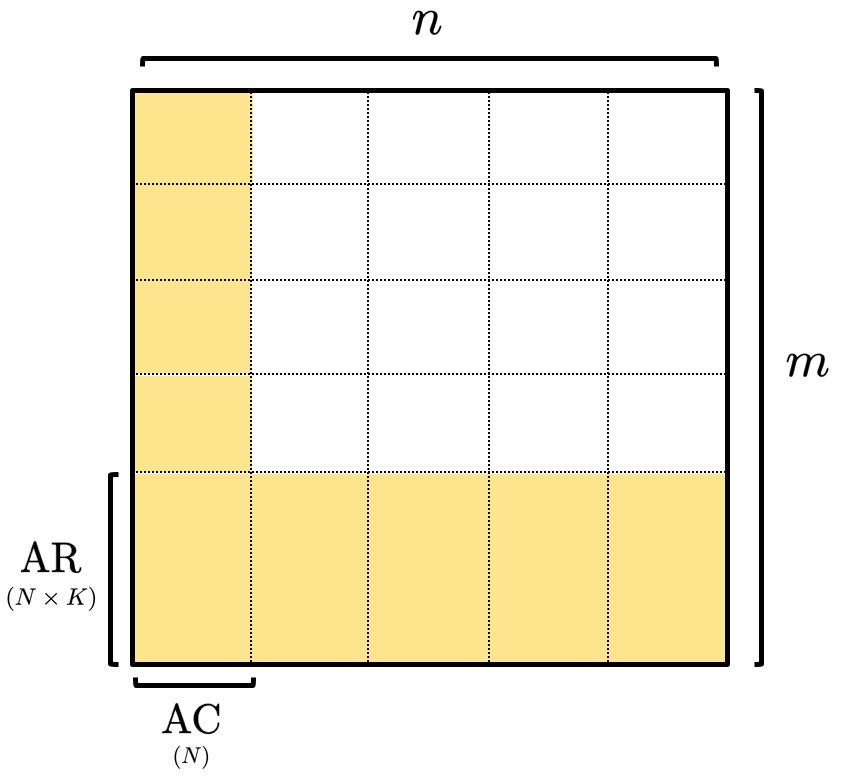}
	\caption{Sparsity pattern for which minimum number of observations required for entry-wise recovery.} 
	\label{fig:optimal_observation_pattern}
\end{figure}

%
Further, in the hypothetical scenario where we get $K$ independent noisy copies for each $(i, j)$, if we wanted to estimate $A_{ij}$ to within error $O(\delta)$ for all $(i, j)$, this would require $m \times n \times K$ observations, with $K = \Omega(\delta^{-2})$.
In contrast, for \SNN, if we assume that for all $(i, j)$, we can use the same set of anchor rows and columns, i.e., $\{|\AC^{(k)}|, |\AR^{(k)}|\}_{k \in [K]}$ can be chosen to be the same for all $(i, j)$, then one can easily verify that the number of observations we need to recover each $A_{ij}$ to within error $O(\delta)$ is at most $N^2 \times K + m \times N + n \times (N \times K)$,\footnote{Technically, we only need $N^2 \times K + (m - N - K) \times N + (n - N) \times (N \times K)$.} with $N = \Omega(\delta^{-1/4})$ and $K = \Omega(\delta^{-2})$.
See Figure \ref{fig:optimal_observation_pattern} for a visual depiction of the observation pattern for which this holds.
Thus, for any fixed $\delta > 0$, we can recover every entry $A_{ij}$ to within additive error $O(\delta)$, with access to only $O(m + n)$ observations, rather than $O(m \times n)$ observations as would be naively required.

{\bf Connections to causal transportability, transfer learning, learning with distribution shift.}
We note that this problem of generalizing well without making an i.i.d assumption is known by a variety of terms across many fields of study; these include ``causal transportability'', ``transfer learning'', ``learning with distribution shift''.
Given that subspace inclusion, i.e., Assumption~\ref{assump:subspace} holds, we show that generalization is possible without making any distributional assumptions about the underlying signal matrix $\bA$.
Indeed, our theoretical results in Theorem~\ref{thm:consistency} and~\ref{thm:normality} can be interpreted as point-wise out-of-sample generalization error bounds, which are distribution free (i.e., instance dependent).
This might be of independent interest.

% section 5: experiments  
\section{Experiments} \label{sec:experiments} 
The objective of this section is to compare the imputation accuracy of \SNN~against the state-of-the-art matrix completion algorithms for MNAR data.
We describe these algorithms in Section \ref{sec:MC_algorithms}.
We do two case studies.
In Section \ref{sec:recsys}, we apply these various algorithms in the setting of recommender systems with different missingness patterns. 
In Section \ref{sec:empirics}, we do the same but using data from a classic panel data case study in the econometrics literature called ``California Prop 99'' \cite{abadie2}.

\subsection{Benchmark Matrix Completion Algorithms}\label{sec:MC_algorithms}
In particular, we compare two types of algorithms for matrix completion against \SNN; we choose these benchmarks to be in line with those considered in \cite{ma2019missing}. 
The first group of algorithms does not account for entries being MNAR; these include \pmf~(\cite{pmf}), \svd~(\cite{funk2006netflix}), \svdpp~(\cite{koren08}), \softimpute~(\cite{softimpute14}), and \knn~(\cite{LeeLiShahSong16}); we remark that the algorithm proposed in \cite{athey2021matrix} is similar to \softimpute~with the addition of separate fixed effects terms. 
In particular, both the algorithm design and associated analysis of these algorithms is for MCAR data. 
In contrast, the second group does account for the limited MNAR setting as described in Section \ref{sec:missingness_models}; these include \maxnorm~(\cite{cai2016matrix}), \expomf~(\cite{Liang2016}), and \wtn~(\cite{WTN}). 
With the exception of \expomf, we further consider IPW-variants of the other benchmark algorithms, i.e., for each algorithm, 
we first de-bias the loss function given in \eqref{eq:erm_weighted} via propensity scores.
We do not do so with \expomf~as their algorithm does not lend itself to be de-based via propensity scores in a straightforward manner (also
see \cite{ma2019missing}).% also do not de-bias \expomf.
The propensity scores are estimated in two ways, which are in line with the MAR and limited MNAR setting described in Section~\ref{sec:missingness_models}.
(i) MAR setting: We provide meaningful additional covariates $(X_i, \tilde{X}_j)$ for row $i$ and column $j$ and use logistic regression to learn $\hp_{ij}$; if a matrix completion algorithm is de-biased in this way, we add \texttt{LR} in front of it, e.g., \texttt{LR-PMF} means the \pmf~algorithm is used to estimate $\bhA$ and the loss function is de-biased using logistic regression.
(ii) Limited MNAR setting: We do not provide additional covariates and directly estimate $\hp_{ij}$ using the observed mask matrix $\bD$; this is done using the \texttt{1bitMC} algorithm in \cite{davenport20141bit} and algorithms de-biased in this manner has a prefix of \texttt{1bitMC} added to them.; this approach to de-bias MNAR data is in line with what is proposed in \cite{ma2019missing, yang2021tenips, bhattacharya2021matrix}.%; 

We consider two error metrics, root mean-squared-error (RMSE) and mean-absolute-error (MAE). 
For all benchmark algorithms, we use 5-fold cross validation to tune their hyper-parameters through grid search for every error metric, i.e., for each benchmark algorithm, we find its best performing hyper-parameters with respect to RMSE and MAE on the validation set and report the error metric-specific hyper-parameters on the test set for each error metric.
For \SNN, we choose $K=1$ and $\lambda^{(1)}$ as per \cite{donoho14}, i.e., we do not tune the hyper-parameters of \SNN~nor do we optimize it for each error metric. 
We emphasize that only the algorithms with the \texttt{LR} prefix use the additional row and column covariates $X_i, \tilde{X}_j$. 

% SIMULATION: 
\subsection{Recommendation Systems} \label{sec:recsys}
We begin with recommendation systems, which is arguably the canonical matrix completion application.
Through the recommendation systems setting, we present two MNAR missingness patterns---one obeys the standard assumptions on MNAR in the literature (which we often refer to as limited MNAR) while the other considers a more general MNAR setting. 
%
%We re-emphasize that positivity underpins the propensity estimation methods used both in the MAR and limited MNAR settings. 
%
To better understand the effect of the underlying mechanism which leads to missingness on each algorithm's ability to perform imputation, we consider the ``noiseless'' case, i.e., $\tY_{ij} = A_{ij}$ if $D_{ij} =1$ and $\tY_{ij} = \star$ otherwise. 
We study the effect of additional noise $\varepsilon_{ij}$ in the panel data setting in Section \ref{sec:empirics}. 

\subsubsection{Limited MNAR Setting: Positivity \& Independent Missingness} \label{sec:standardmnar} 
In our first illustration, our observation pattern reflects the self-selection bias phenomena where most users tend to provide ratings if they particularly liked or disliked an item. 
However, they are much less inclined to provide a rating for an item that they are lukewarm about. 
Our simulated setup also consists of ``core users'' and ``core movies''. 
We use core users to represent movie fanatics or critics, for instance, who provide explicit feedback for a significant number of films.
%; additionally, due to their social influence, their opinions often shape those within their social network. 
% 
In the setting of movie recommendations, we use core items to represent iconic movies such as {\em Star Wars} or {\em Titanic} that have influenced future films and popular culture, and are largely viewed by the general audience. 
These can also represent the subset of items that online platforms such as {\em Hulu, Netflix, Stichfix} display to new users when prompting for their preferences.

{\bf Experimental setup.}
We consider $m = 80$ users and $n=80$ movies. 
We choose the dimension of the latent space as $r=5$. 
We generate the latent user matrix $\bU \in \Rb^{m \times r}$ as follows: 
(i) we first choose $m_\core =20$ core users and construct $\bU_0 \in \Rb^{m_\core \times r}$ by sampling entries i.i.d. from a standard normal distribution; 
(ii) next, we construct $\bU_1 = \bB \bU_0 \in \Rb^{(m-m_\core) \times r}$, where the entries in $\bB \in \Rb^{(m-m_\core) \times m_\core}$ are sampled i.i.d. from a Dirichlet distribution, which ensures that the new factors lie in the same intervals as the factors in $\bU_0$. 
In doing so, every row of $\bU_1$, representing the latent factors corresponding to the ``standard'' users, is a linear combination of that of core users $\bU_0$, i.e., every standard user can be expressed as a weighted combination of core users. 
We then define $\bU = [\bU_0, \bU_1]$ such that the first $m_\core$ rows of $\bU$ correspond to the core users. 
We construct $\bV = [\bV_0, \bV_1] \in \Rb^{n \times r}$ similarly, where $\bV_0 \in \Rb^{n_\core \times r}$ and $\bV_1 \in \Rb^{(n - n_\core) \times r}$ represent the matrix of latent factors associated with core movies and standard movies, respectively; 
here, we choose $n_\core = 20$. 
We form $\bA = \bU \bV^T \in \Rb^{m \times n}$ and scale the values to lie within the interval $[1, 5]$; by construction, $\bA$ is a low-rank matrix.

Finally, we generate user and movie covariates matrices $\bX = \bU \bQ_1 \in \Rb^{m \times 3}$ and $\tilde{\bX} = \bV \bQ_2 \in \Rb^{n \times 3}$, where the entries in $\bQ_1 \in \Rb^{r \times 3}$ and $\bQ_2 \in \Rb^{r \times 3}$ sampled i.i.d. from a standard normal $\mathcal{N}(0,1)$; 
additionally, we normalize the columns in $\bQ_1$ and $\bQ_2$ to have unit $\ell_2$-norm. 

Next, we describe our generative model for the propensity matrix $\bP \in \Rb^{m \times n}$. 
Without loss of generality, we denote
$\Cc_\core := \{(i,j): i \le m_\core, j \le n_\core\}$ as the subset of core users and core movies, 
$\Cc_\tuser := \{(i,j): i \le m_\core, j > n_\core\}$ as the subset of core users and standard movies,
$\Cc_\titem := \{(i,j): i > m_\core, j \le n_\core\}$ as the subset of standard users and core movies, and
$\Cc_\standard := \{(i,j): i > m_\core, j > n_\core\}$ as the subset of standard users and standard movies. 
These will represent our four cohorts of interest. 
Next, for some $t \in (1, 5)$, $\kappa_{ij} > 0$, and $\alpha_{ij} \in (0,1]$, 
\begin{align}
	p_{ij} = \begin{cases}
		& \kappa_{ij} \cdot \alpha_{ij}^{A_{ij}-1}, \quad \text{if } A_{ij} \in [1, t]
		\\ & \kappa_{ij} \cdot \alpha_{ij}^{5-A_{ij}}, \quad \text{if } A_{ij} \in (t, 5]. 
	\end{cases}
\end{align}
In our setting, we choose our threshold $t = 2.3$. 
Here, $\alpha_{ij}$ is a parameter that controls the MNAR effect: $\alpha_{ij} = 1$ is MCAR while $\alpha_{ij} \rightarrow 0$ only reveals $1$ and $5$ rated movies. 
We choose $\alpha_{ij} = 0.7$ for $(i,j) \in \Cc_\core$, 
$\alpha_{ij} = 0.35$ for $(i,j) \in \Cc_\tuser, \Cc_\titem$,
and $\alpha_{ij}=0.1$ for $(i,j) \in \Cc_\standard$. 
For every $(i,j)$ pair, $\kappa_{ij}$ is set so that the expected number of revealed ratings within the cohort is equal to some value. 
We choose the expected number of observations within $\Cc_\core$ as 90\%, within $\Cc_\tuser$ as 70\%, within $\Cc_\titem$ as 70\%, and within $\Cc_\standard$ as 5\%. 
This sampling process ensures the two key assumptions in the limited MNAR setting of the entries of $\bD$ being independent and $p_\min > 0$ are satisfied.
See Figure \ref{fig:sparsity_limited_MNAR} for a visual depiction of empirical sparsity pattern under this missingness mechanism.

{\bf Results.}
In the following simulations, we obey the generative process above.
In particular, we sample $\bA$ and $\bP$ once, as well as $\bX$ and $\tilde{\bX}$, and perform 10 experimental repeats where the only randomization lies in the sparsity pattern, i.e., we observe 10 independent realizations of $\bD$. 
We report the average RMSEs and MAEs, as well as their respective standard deviations, over the 10 experimental runs in Table~\ref{table:experiments}.
We find that with respect to MAE, \SNN~achieves the best result along with \maxnorm~(and its variants);
with respect to RMSE, \SNN~is a close second with \svdpp~(and its variants), after \maxnorm~(and its variants). 
Although positivity and independence between entries in $\bD$ are upheld, we remark that debiasing via \texttt{1bitMC} and \texttt{LR-} do not always yield stronger results, e.g., see \pmf~and \softimpute.

% SIMULATION: STAIRCASE
\subsubsection{A More General MNAR Setting: Violating Positivity \& Independence Assumptions}
\label{sec:generalmnar}
In this simulation, we violate two key assumptions in the current literature on MNAR data: (i) positivity and (ii) independence between the entries in $\bD$. 
Towards this, we continue the notion of core movies, for which all users provide ratings. 
For the remaining movies, users only provide ratings if a movie belong to their favorite genre. 
This deterministically sets every entry in $\bP$ (and thus $\bD$) to either $0$ or $1$, and correlates the entries in $\bD$, which yields a sparsity pattern similar to that shown in Figure~\ref{fig:staircase}. 
See Figure \ref{fig:sparsity_general_MNAR} for a visual depiction of empirical sparsity pattern under this missingness mechanism.

{\bf Experimental setup.}
In particular, we consider $m=80$ users and $n=80$ items. 
We choose the dimension of the latent space as $r=5$. 
We generate the latent user matrix $\bU \in \Rb^{m \times r}$ by sampling entries i.i.d. from a standard normal distribution. 
To generate the latent item matrix $\bV \in \Rb^{n \times r}$, we first choose $n_{\text{core}}=30$ core items (to be defined in greater detail below) and construct $\bV_0 \in \Rb^{n_{\text{core}} \times r}$ by sampling entries i.i.d. from a standard normal. 
Next, we construct $\bV_1 = \bB \bV_0 \in \Rb^{(n-n_{\text{core}}) \times r}$, where the entries in $\bB \in \Rb^{(n-n_{\text{core}}) \times n_{\text{core}}}$ are sampled i.i.d. from a Dirichlet distribution. 
In doing so, every row of $\bV_1$ is a linear combination of rows in $\bV_0$, i.e., every item can be expressed as a weighted combination of core items. 
We then define $\bV = [\bV_0, \bV_1]$ such that the first $n_{\text{core}}$ rows of $\bV$ correspond to the core items.
We form $\bA = \bU \bV^T \in \Rb^{m \times n}$ and scale the values to lie within the interval $[1,5]$. 
Finally, we generate user and item feature matrices $\bX = \bU \bQ_1 \in \Rb^{m \times 10}$ and $\tilde{\bX} = \bV \bQ_2 \in \Rb^{n \times 10}$, where the entries in $\bQ_1 \in \Rb^{r \times 10}$ and $\bQ_2 \in \Rb^{r \times 10}$ sampled i.i.d. from a standard normal $\mathcal{N}(0,1)$; 
additionally, we normalize the columns in $\bQ_1$ and $\bQ_2$ to have unit $\ell_2$-norm. 
We generate higher dimensional covariates $\bX$ and $\tilde{\bX}$ to see if improves the relative performance of the MAR algorithms, denoted by the prefix \texttt{LR}, which use this additional information to estimate the propensities.   

To describe the generating process for the observation pattern $\bD$, we begin by providing an interpretation of the above quantities.  
First, we interpret $r$ as the number of latent genres.
In turn, the $(i,k)$-th entry in $\bU$ can be interpreted as user $i$'s preference for genre $k$;
similarly, the $(j, k)$-th entry in $\bV$ can be interpreted as the level to which item $j$ is composed of genre $k$.
We consider the setting where all users provide ratings for all core items, i.e., $D_{ij} = 1$ for every user $i \in [m]$ and core item $j \in [n_{\text{core}}]$.
%
%As discussed in Section~\ref{sec:alg_anchor_applications}, this is a caricature for online platforms such as {\em Hulu} that ask new users to indicate their preferences from a small subset of movies at the commencement of their subscription, and iconic movies such as {\em Star Wars} or {\em Titanic} that have largely influenced future films and popular culture for which most users have watched. 
%
For the remaining entries in $\bD$, we posit that every user will only rate items from their favorite genre. 
More specifically, given the above interpretation, we define user $i$'s favorite genre $k^*(i)$ as $k^*(i) = \argmax_{k \in [r]} U_{ik}$;
similarly, we classify an item $j$ as belonging to genre $k^\sharp(j)$ if $k^\sharp(j) = \argmax_{k \in [r]} V_{jk}$. 
Hence, for every user $i \in [m]$ and non-core item $j > n_{\text{core}}$, we have $D_{ij} = 1$ if $k^*(i) = k^\sharp(j)$ and $0$ otherwise. 
We underscore that this model violates the standard operating assumptions within the current MNAR literature as entries in $\bD$ are deterministically set to $0$ (i.e., the minimum element in $\bP$ is $0$), and are dependent on one another. 
%
%For every pair $(i,j)$ satisfying $D_{ij}=1$, we observe a discretized version of $A_{ij}$. 
%%
%The discretization is carried out as follows: 
%%
%Let $\delta_{ij}$ be the decimal point of $A_{ij}$, e.g., if $A_{ij} = 2.7$, then $\delta_{ij} = 0.7$. 
%%
%We define our observation $\tY_{ij}$ as $\tY_{ij} = \ceil{A_{ij}}$ w.p. $\delta_{ij}$ and $\tY_{ij} = \floor{A_{ij}}$ w.p. $1-\delta_{ij}$, 
%e.g., $\tY_{ij} = 3$ w.p. $0.7$ and $\tY_{ij} = 2$ w.p. $0.3$. 
%

{\bf Results.}
In the following simulations, we obey the generative process above. 
In particular, we sample $\bV$ and $\bX$ once, and perform 10 experimental repeats where the only randomization lies in the re-sampling of $\bU$; 
this is done to model new users coming into the system with the movies fixed. 
We report the average RMSEs and MAEs, as well as their respective standard deviations, over the 10 experimental runs in Table~\ref{table:experiments}.
We find that \SNN~achieves the best RMSE and MAE, with \texttt{1bitMC}-\maxnorm~as a close second with respect to RMSE and MAE. 
As with the limited MNAR setting experiment, we find that de-biasing does not always improve results. 
This is reasonable given that our generative process violates the typical assumptions underlying propensity estimation methods. 
The relative improvement of \SNN~shows its robustness to the general MNAR setting, where entry-wise positivity and  independence of $\bD$ are violated.
The fact that $\knn$~performs relatively poorly indicates that matching via linear weights is indeed more expressive than matching with uniforms weights as done in $\knn$.
WE also note that the various state-of-the-art algorithms are still relatively robust to the general MNAR setting.
This may warrant further investigation into the potential gap between theory and practice on the robustness of these methods to different missingness patterns.

%%
%As discussed in Section~\ref{sec:snn_alg}, \SNN~shares similarities with \knn; thus, it is unsurprising that both algorithms achieve similar results.
%%
%Nevertheless, \SNN~still outperforms \knn, which is to be expected given that the non-core items are formed as a weighted combinations of core items. 
%%
%More surprising, however, is the empirical success of \svd~and \svdpp~despite the violation of standard operating assumptions.
%%
%Therefore, it may be of independent interest to study the robustness properties of both algorithms given their potential gap between theory and practice.
%%
%Finally, we remark that the debiasing techniques via \texttt{1bitMC} and \texttt{LR-} do not always yield stronger results.
%%
%This is reasonable given that our generative process violates the typical assumptions underlying propensity estimation methods. 

\begin{table}[t!]
\centering 
\smaller 
\begin{tabular}{@{}ccccccc@{}}
\toprule
\multirow{2}{*}{Algorithm} & \multicolumn{2}{c}{Rec Sys (limited MNAR)}  & \multicolumn{2}{c}{Rec Sys (general MNAR)} & \multicolumn{2}{c}{Panel Data}   \\ \cmidrule(l){2-7} 
		        & RMSE   	     		& MAE		        & RMSE   	     	& MAE	& RMSE 	&MAE	\\
		 \midrule 
\pmf     		& $0.30 \pm 0.02$	& $0.25 \pm 0.02$   & $0.64 \pm 0.14$	    & $0.56 \pm 0.11$		& $89.4 \pm 92$	    & $105 \pm 94$    \\
\bitpmf     		& $0.42 \pm 0.02$	& $0.36 \pm 0.02$   & $0.69 \pm 0.12$	    & $0.61 \pm 0.11$	 	& $69.2 \pm 85$	    & $84.6 \pm 91$   \\
\lrpmf     		& $0.28 \pm 0.02$	& $0.28 \pm 0.02$   & $0.66 \pm 0.15$	    & $0.57 \pm 0.13$	  	& $32.1 \pm 56$	    & $47.5 \pm 76$	  \\
\midrule

\svd    		& $0.14 \pm 0.01$	& $0.11 \pm 0.00$	& $0.49 \pm 0.03$	& $0.39 \pm 0.03$		& $14.9 \pm 2.3$	& $10.9 \pm 1.6$	\\
\bitsvd    		& $0.15 \pm 0.01$	& $0.12 \pm 0.01$	& $0.49 \pm 0.03$	& $0.39 \pm 0.03$		& $15.0 \pm 2.3$	& $10.9 \pm 1.6$	\\
\lrsvd    		& $0.14 \pm 0.01$	& $0.11 \pm 0.00$	& $0.49 \pm 0.03$	& $0.39 \pm 0.03$		& $15.0 \pm 2.3$	& $10.9 \pm 1.6$	\\
\midrule

\svdpp     		& $0.07 \pm 0.02$	& $0.06 \pm 0.01$	& $0.44 \pm 0.03$	    & $0.34 \pm 0.03 $	  	& $161 \pm 76$	    & $160 \pm 76$  \\
\bitsvdpp  		& $0.08 \pm 0.02$	& $0.08 \pm 0.01$	& $0.45 \pm 0.03$	    & $0.35 \pm 0.03$	   	& $143 \pm 86$	    & $141 \pm 87$ \\
\lrsvdpp    		& $0.08 \pm 0.02$	& $0.08 \pm 0.01$	& $0.44 \pm 0.03$	    & $0.35 \pm 0.03$	   	& $180 \pm 57$	    & $178 \pm 57$ \\
\midrule

\softimpute     	& $1.03 \pm 0.05$	& $0.89 \pm 0.04$	& $1.50 \pm 0.06$	    & $1.44 \pm 0.05$		& $101 \pm 4.1$	    & $99.1 \pm 4.1$	\\
\bitsoftimpute  	& $1.21 \pm 0.04$	& $1.06 \pm 0.03$	& $1.52 \pm 0.09$	    & $1.46 \pm 0.09$		& $100 \pm 4.1$	    & $97.7 \pm 4.1$	\\
\lrsoftimpute   	& $1.03 \pm 0.05$	& $0.89 \pm 0.04$	& $1.50 \pm 0.06$	    & $1.44 \pm 0.05$		 & $103 \pm 3.8$	& $101 \pm 3.9$   \\
\midrule

\wtn     		& $0.13 \pm 0.01$	& $0.10 \pm 0.01$	& $0.52 \pm 0.13$		& $0.44 \pm 0.11$		& $99.9 \pm 4.1$	& $97.7 \pm 4.1$\\
\bitwtn     		& $0.10 \pm 0.01$	& $0.08 \pm 0.00$	& $0.55 \pm 0.15$	   	& $0.47 \pm 0.14$		& $100 \pm 4.1$	    & $97.8 \pm 4.1$\\
\lrwtn    		& $0.12 \pm 0.01$	& $0.10 \pm 0.00$	& $0.52 \pm 0.16$		& $0.43 \pm 0.15$		& $99.9 \pm 4.1$	& $97.8 \pm 4.1$\\
\midrule

\maxnorm     	& $0.05 \pm 0.01$	& $0.03 \pm 0.01$	& $0.29 \pm 0.08$		& $0.20 \pm 0.06$		& $99.9 \pm 4.1$	& $97.7 \pm 4.1$\\
\bitmaxnorm     	& $0.05 \pm 0.01$	& $0.03 \pm 0.01$	& $0.23 \pm 0.06$	   	& $0.17 \pm 0.05$		& $100 \pm 4.1$	    & $97.8 \pm 4.1$\\
\lrmaxnorm     	& $0.05 \pm 0.01$	& $0.03 \pm 0.01$	& $0.31 \pm 0.09$	    	& $0.20 \pm 0.06$		& $100 \pm 4.1$	    & $97.8 \pm 4.1$\\
\midrule

\expomf     	& $2.08 \pm 0.01$	& $2.00 \pm 0.01$	& $1.99 \pm 0.05$		& $1.90 \pm 0.05$		& $75.0 \pm 5.5$	& $64.8 \pm 19$\\
\midrule

\knn     		& $0.51 \pm 0.02$	& $0.40 \pm 0.02$	& $0.40 \pm 0.06$		& $0.30 \pm 0.05$		& $15.4 \pm 2.5$	& $12.0 \pm 1.7$\\
\midrule

\SNN     		& $0.08 \pm 0.01$	& $0.03 \pm 0.01$	&  $0.20 \pm 0.07$		& $0.11 \pm 0.06$		&  $10.3 \pm 1.0$	& $8.00 \pm 0.7$\\
\bottomrule
\end{tabular}
\caption{\smaller RMSEs and MAEs of matrix completion methods on a recommender system experiment and a panel data experiment.
The first two columns correspond to Section~\ref{sec:standardmnar}, 
the middle two columns correspond to Section~\ref{sec:generalmnar},
and the final two columns correspond to Section~\ref{sec:empirics}. 
The results are the averages $\pm$ standard deviations across 10 experimental repeats.}
\label{table:experiments}
\end{table}

\subsection{Panel Data} \label{sec:empirics} 
We now compare \SNN~against the same benchmark matrix completion algorithms using a classic case study of California smoking data of \cite{abadie2}, which has been widely utilized within the econometrics literature. 
We do so as this setting has a MNAR sparsity pattern which is quite distinct from what is seen in recommendation systems.
We now give a brief overview of the case study.
In 1988, California introduced the 
first modern-time large-scale anti-tobacco legislation in the United States (Proposition 99). 
There was interest in estimating the effect of this legislation on tobacco sales in California.
Towards this, per-capita cigarette sales data was collected across 39 U.S. states from 1970 to 2000. 
Among the 39 states, there was one ``treated'' state, California, which implemented the legislation; the remaining 38 states were chosen as ``control'' states as they neither instituted a tobacco control program nor raised cigarette sales taxes by 50 cents or more. 
These other 38 control states were then used to build a ``synthetic California'', i.e., a synthetic trajectory of cigarette sales in California if it had not introduced any tobacco legislation.

{\bf Experimental setup.}\label{sec:experimental_setup} 
We consider the time horizon of $n=31$ years and restrict our focus to the $m = 38$ control units in the original dataset.
%i.e., states that did not adopt a large-scale tobacco control program or raise state cigarette taxes by $50$ cents or more during the years 1989 to 2000. 
%
This data is encoded into a $38 \times 31$ matrix, $\bY$, where the entry $Y_{ij}$ represents the potential outcome of per-capita cigarette sales (in packs) for state $i$ in year $j$ under control, i.e., without any intervention in place. 
To generate MNAR data, we artificially introduce interventions to a subset of states in 1989, where the probability a state adopts an intervention (e.g., tobacco control program) depends on their change in cigarette sales pre- and post-1989. 
More specifically, we consider the following treatment adoption protocol: 
First, we cluster states into three categories---mild, moderate, or severe---based on their change in average cigarette sales during 1989-2000 compared to that during 1970-1988; we note that in this context, a negative change means that the cigarette sales in the post-intervention period are lower than that in the pre-intervention period. 
As such, we define 
(i) mild states as those whose change is at least one standard deviation above the average change across all states; 
(ii) severe states as those whose change is at least one standard deviation below the average change across all states;
(iii) and moderate states as the remaining states whose change is within one standard deviation. 

We then designate the probability of intervention for mild, moderate, and severe states as $10\%$, $30\%$, and $50\%$, respectively. 
In words, this setup reflects the scenario in which a state is more likely to adopt an intervention if their average sales in the post-intervention period is relatively closer to their pre-intervention sales compared to that of their peer states.  
In the language of causal inference, this is exactly confounding, i.e., there is a correlation between the treatment assignment and the eventual outcome.

\begin{figure}[t!]
	\centering 
	\begin{subfigure}[b]{0.3\textwidth}
		\centering 
		\includegraphics[width=\linewidth]
		{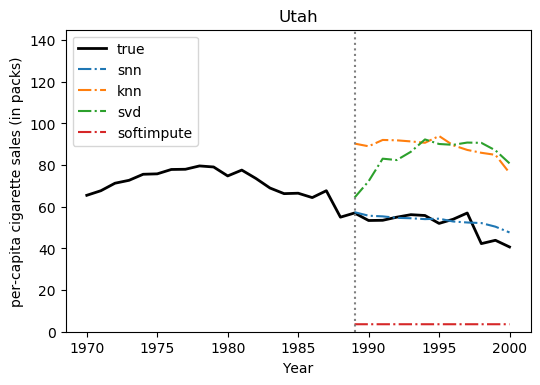}
		\caption{\smaller Mild state: Utah.} 
		\label{fig:s1_new} 
	\end{subfigure} 
	\hfill 
	\begin{subfigure}[b]{0.3\textwidth}
		\centering 
		\includegraphics[width=\linewidth]
		{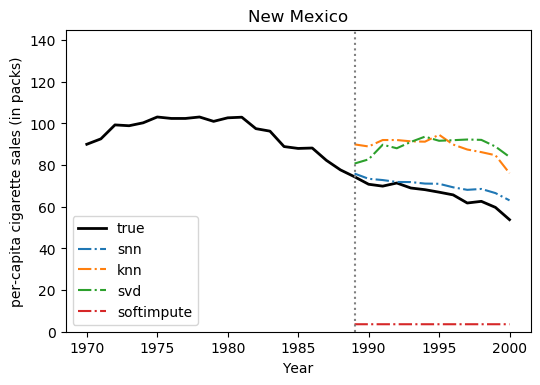}
		\caption{\smaller Moderate state: New Mexico.} 
		\label{fig:s2_new}
	\end{subfigure} 
	\hfill 
	\begin{subfigure}[b]{0.3\textwidth}
		\centering 
		\includegraphics[width=\linewidth]
		{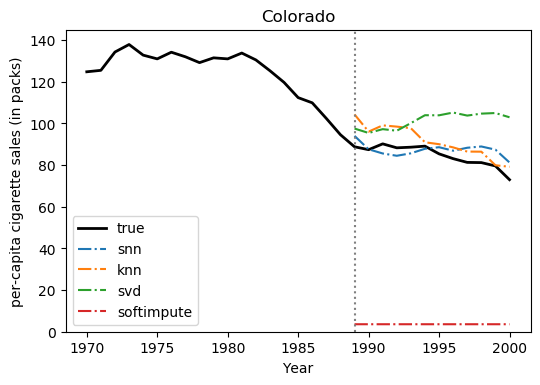}
		\caption{\smaller Severe state: Colorado.} 
		\label{fig:s3_new}
	\end{subfigure} 
	\caption{\smaller True observations are represented in black, \SNN~estimates shown in blue,  \knn~estimates shown in orange, \svd~estimates shown in green, \softimpute~estimates shown in red. 
	}
	\label{fig:prop99states} 
\end{figure}

For an example of a mild, moderate, and severe state, please see Figure~\ref{fig:prop99states}. 
Additionally, we remark that once an intervention is adopted, all sales under control during the post-intervention period are, by definition, unobserved, i.e., $\tY_{ij} = \star$ for any  intervened on state $i$ and for all $j \ge 19$ (after 1988); this yields the observation pattern shown in Figure~\ref{fig:panel}.
Finally, to employ logistic regression, i.e., \texttt{LR} to de-bias the estimates, we use state covariate data from \cite{abadie2}, which include average retail price of cigarettes, per capita state personal income (logged), the percentage of the population age 15-24, and per capita beer consumption.  
We note that \SNN~does not use this auxiliary data.

{\bf Results.} 
Using the above setup, we apply the various matrix completion methods to impute the missing counterfactual cigarette sales associated with the artificial intervention states during the post-intervention period. 
We report the average root mean-squared-errors (RMSEs) and mean absolute errors (MAEs), as well as their respective standard deviations, over 10 experimental runs in Table~\ref{table:experiments}. 
As the table shows, \SNN~significantly outperforms all baseline algorithms under both error metrics. 
The only exception is \knn, which performs similarly to \SNN; this is interesting as \knn~is in essence, the difference-in-differences estimator, a standard method within the panel data econometrics literature.
Further, \svdpp~and \maxnorm~(and its variants), which performed strongly in the recommendation systems example, now incur a significant error.
We display a few representative results in Figure~\ref{fig:prop99states}. 
Collectively across all three studies, we find that \SNN~is robust under varying missingness mechanisms. 

\bibliographystyle{apalike}
\bibliography{bib}

\newpage
\appendix

\section{Original \USVT~Algorithm Experiments}\label{teaser:standard_USVT}
We run the same experiments in Section \ref{sec:intro_teaser} using the original \USVT~estimator of \cite{Chatterjee15} rather than the modified version as proposed in \cite{bhattacharya2021matrix} for MNAR data.
See Figure \ref{fig:standard_USVT} below.
Interestingly, we find the original \USVT~estimator performs better.
Compare Figures \ref{fig:standard_USVT_MCAR}, \ref{fig:standard_USVT_limited_MNAR}, 
\ref{fig:standard_USVT_general_MNAR} with Figures \ref{fig:teaser_MCAR_USVT}, \ref{fig:teaser_limited_MNAR_USVT}  \ref{fig:teaser_general_MNAR_USVT}, respectively.

\begin{figure}[h!]
	\centering 
	\begin{subfigure}[b]{0.3\textwidth}
		\centering 
		\includegraphics[width=\linewidth]
		{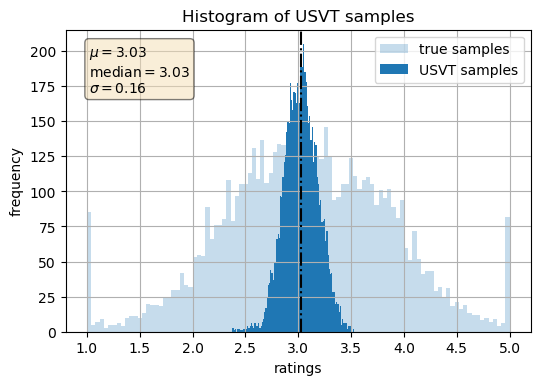}
		\caption{\smaller MCAR.} 
		\label{fig:standard_USVT_MCAR}
	\end{subfigure} 
	\qquad	\begin{subfigure}[b]{0.3\textwidth}
		\centering 
		\includegraphics[width=\linewidth]
		{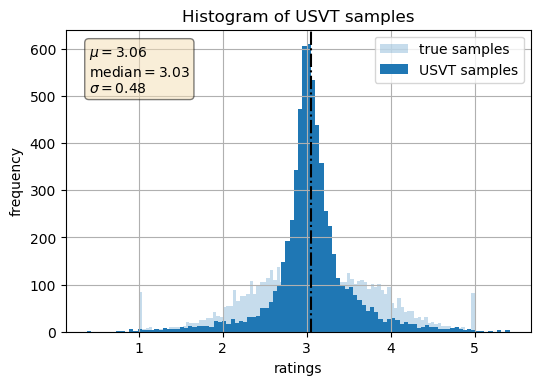}
		\caption{\smaller Limited MNAR.} 
		\label{fig:standard_USVT_limited_MNAR} 
	\end{subfigure}
	\qquad 
	\begin{subfigure}[b]{0.3\textwidth}
		\centering 
		\includegraphics[width=\linewidth]
		{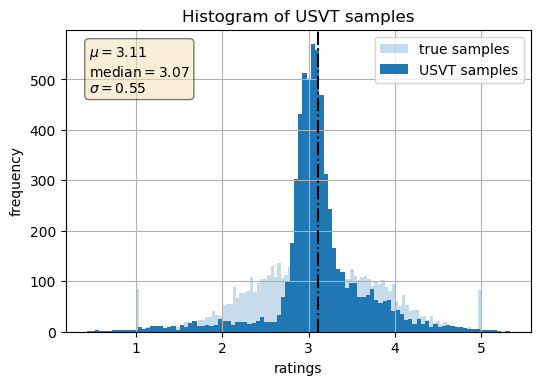}
		\caption{\smaller General MNAR.} 
		\label{fig:standard_USVT_general_MNAR} 
	\end{subfigure} 
	\caption{\smaller Original \USVT~algorithm under the three different experiments.}
	\label{fig:standard_USVT} 
\end{figure} 
\section{Proof of Theorem~\ref{thm:identification}}

In what follows, the descriptors above the equalities represent the assumption used, e.g., $A1$ represents Assumption 1: 
\begin{align} 
	A_{ij} &= \Ex[Y_{ij} | u_i, v_j]
	\\ &\stackrel{A1}= \Ex[\langle u_i, v_j \rangle + \varepsilon_{ij} ~|~ u_i, v_j] 
	\\ &\stackrel{A2}= \langle u_i, v_j \rangle  ~|~ \{u_i, v_j\} 
	\\ &
	= \langle u_i, v_j \rangle  ~|~\bU, \bV, \bD 
	\\ &\stackrel{A3}=  \sum_{\ell \in \Ic} \beta_\ell \cdot \langle u_\ell,  v_j \rangle  ~|~\bU, \bV, \bD
	\\ &\stackrel{A2}=  \sum_{\ell \in \Ic} \beta_\ell \cdot \Ex\left[ \langle u_\ell, v_j \rangle + \varepsilon_{\ell j} ~|~ \bU, \bV, \bD \right] 
	\\ &\stackrel{A1}= \sum_{\ell \in \Ic} \beta_\ell \cdot \Ex\left[ Y_{\ell j} |\bU, \bV, \bD\right]
	\\ &= \sum_{\ell \in \Ic} \beta_\ell \cdot \Ex\left[ \tY_{\ell j} |\bU, \bV, \bD\right]. 
\end{align} 
  
\section{Proof of Theorem \ref{thm:consistency}}
%
%\paragraph{Notation.} 
For ease of notation, we suppress the conditioning on $\Ec$ for the remainder of the proof.
Further, for every $k$, let $\varepsilon^{(k)} = [\varepsilon_{\ell j}: \ell \in \AR^{(k)} ] \in \Rb^{|\AR^{(k)}|}$ and $\Delta^{(k)} = \hbeta^{(k)} - \tbeta^{(k)}$. 
We also recall the definitions provided in Section~\ref{sec:estimator}. 
%Further, for any matrix $\bX$ with orthonormal columns, let $\Pc_X = \bX \bX^T$ denote the projection matrix onto the subspace spanned by the columns of $\bX$. 

To begin, recall that $|\AR^{(k)}| \ge \mu$ for each $k$ by assumption. Thus, by Theorem~\ref{thm:identification}, there exists a $\beta^{(k)} \in \Rb^{|\AR^{(k)}|}$ such that $A_{ij} = \langle \Ex[x^{(k)}], \beta^{(k)} \rangle$ for every $k$, i.e., 
\begin{align} \label{eq:00}
	A_{ij} = \frac{1}{K} \sum_{k=1}^K \langle \Ex[x^{(k)}], \beta^{(k)} \rangle. 
\end{align} 
Additionally, under Assumption~\ref{assump:subspace}, we have $\Ex[x^{(k)}] = \Pc_{U^{(k)}} \Ex[x^{(k)}]$. In turn, this implies
\begin{align} 
	\langle \Ex[x^{(k)}], \beta^{(k)} \rangle &= \langle \Ex[x^{(k)}], \tbeta^{(k)} \rangle \label{eq:0.a} 
	\\ \langle \Ex[x^{(k)}], \Delta^{(k)} \rangle &= \langle \Ex[x^{(k)}], \Pc_{U^{(k)}} \Delta^{(k)} \rangle, \label{eq:0.b} 
\end{align} 
where $\tbeta^{(k)} = \Pc_{U^{(k)}} \beta^{(k)}$. 
Together, \eqref{eq:00}, \eqref{eq:0.a}, and \eqref{eq:0.b} yield the following:
\begin{align}
	\hA_{ij} - A_{ij} &= \frac{1}{K} \sum_{k=1}^K \left( \hA_{ij}^{(k)} - A_{ij} \right) 
	\\ &= \frac{1}{K} \sum_{k=1}^K \left( \langle x^{(k)}, \hbeta^{(k)} \rangle - \langle \Ex[x^{(k)}], \beta^{(k)} \rangle \right) 
	\\ &= \frac{1}{K} \sum_{k=1}^K \left( \langle x^{(k)}, \hbeta^{(k)} \rangle - \langle \Ex[x^{(k)}], \tbeta^{(k)} \rangle \right) 
	\\ &= \frac{1}{K} \sum_{k=1}^K \left( \langle \Ex[x^{(k)}], \Delta^{(k)} \rangle 
	+ \langle \varepsilon^{(k)}, \tbeta^{(k)} \rangle + \langle \varepsilon^{(k)}, \Delta^{(k)} \rangle \right)
	\\ &= \frac{1}{K} \sum_{k=1}^K \left( \langle \Ex[x^{(k)}], \Pc_{U^{(k)}} \Delta^{(k)} \rangle 
	+ \langle \varepsilon^{(k)}, \tbeta^{(k)} \rangle + \langle \varepsilon^{(k)}, \Delta^{(k)} \rangle \right). \label{eq:1} 
\end{align}
Below, we bound the three terms on the right-hand side (RHS) of \eqref{eq:1} separately. 

% term 1
{\em Bounding term 1.} 
By Cauchy-Schwartz inequality, we obtain for every $k$ 
\begin{align}
	\langle \Ex[x^{(k)}], \Pc_{U^{(k)}} \Delta^{(k)} \rangle \le \| \Ex[x^{(k)}] \|_2 \cdot \|  \Pc_{U^{(k)}} \Delta^{(k)} \|_2. 
\end{align}
Under Assumption~\ref{assump:bounded}, we have $ \| \Ex[x^{(k)}] \|_2 \le |\AR^{(k)}|^{1/2}$. 
As such,
\begin{align} \label{eq:2} 
	\frac{1}{K} \sum_{k=1}^K \langle \Ex[x^{(k)}], \Pc_{U^{(k)}} \Delta^{(k)} \rangle 
	&\le \frac{1}{K} \sum_{k=1}^K |\AR^{(k)}|^{1/2} \cdot \|  \Pc_{U^{(k)}} \Delta^{(k)} \|_2. 
	%\\ &\le \max_{k} ~|\AR^{(k)}|^{1/2} \cdot \|  \Pc_{U^{(k)}} \Delta^{(k)} \|_2. 
\end{align}  
To bound the expression above, we use the following lemma; its proof is found in Appendix~\ref{sec:lemma_proof_1}. 
%Lemma F.1 of \cite{agarwal2021synthetic}. 
%
\begin{lemma}[Lemma G.1 of \cite{agarwal2021synthetic}] \label{lemma:param_est}
Consider the setup of Theorem~\ref{thm:consistency}. Then for any $k$,
\begin{align}
	 \Pc_{U^{(k)}} \Delta^{(k)} &= O_p \left( \frac{(r^{(k)})^{1/2}}{|\AR^{(k)}|^{1/2} |\AC^{(k)}|^{1/4}} 
	 + \frac{ (r^{(k)})^{3/2} \| \tbeta^{(k)}\|_1 \log^{1/2}(|\AC^{(k)}| |\AR^{(k)}|)} {|\AR^{(k)}|^{1/2} ~ \min\{ |\AC^{(k)}|^{1/2}, |\AR^{(k)}|^{1/2} \}}\right). 
\end{align} 
\end{lemma} 
Plugging Lemma~\ref{lemma:param_est} into \eqref{eq:2}, we conclude
\begin{align} \label{eq:lemma.1}
	\frac{1}{K} \sum_{k=1}^K \langle \Ex[x^{(k)}], \Pc_{U^{(k)}} \Delta^{(k)} \rangle &= 
	O_p\left( 
	% + \max_k
	 \frac{1}{K}\sum_{k=1}^K \frac{(r^{(k)})^{1/2}}{ |\AC^{(k)}|^{1/4}}  
	 + \frac{(r^{(k)})^{3/2} \| \tbeta^{(k)}\|_1 \log^{1/2}(|\AC^{(k)}| |\AR^{(k)}|)} { \min\{ |\AC^{(k)}|^{1/2}, |\AR^{(k)}|^{1/2} \}}\right).
\end{align} 

% term 2
{\em Bounding term 2.} 
We begin with a lemma that is an immediate consequence of Hoeffding's Lemma. 
\begin{lemma} \label{lemma:a_hoeffding}
Let $\gamma_k$ be a sequence of mean zero sub-gaussian r.v.s with $\Ex[\gamma_k^2] = \sigma_k^2$. 
Then, 
\begin{align}
	\frac{1}{K} \sum_{k=1}^K \gamma_k = O_p \left( \frac{1}{K} \left[ \sum_{k=1}^K \sigma^2_k \right]^{1/2} \right). 
\end{align}
\end{lemma}
By Assumption~\ref{assump:subg}, we have for any $k$, 
\begin{align}
	\Ex[\langle \varepsilon^{(k)}, \tbeta^{(k)}\rangle] &= 0
	\\ \text{Var}(\langle \varepsilon^{(k)}, \tbeta^{(k)}\rangle) &= \sum_{\ell \in \AR^{(k)}} (\tbeta^{(k)}_\ell \sigma_{\ell j})^2. \label{eq:variance_per_k}
\end{align}
Since $\langle \varepsilon^{(k)}, \tbeta^{(k)}\rangle$ are independent across $k$, noting that $\sum_{\ell \in \AR^{(k)}} (\tbeta^{(k)}_\ell \sigma_{\ell j})^2 \le \sigma^2 \| \tbeta^{(k)} \|_2^2$, and applying Lemma~\ref{lemma:a_hoeffding} yields 
\begin{align} \label{eq:lemma.2}
	\frac{1}{K} \sum_{k=1}^K \langle \varepsilon^{(k)}, \tbeta^{(k)}\rangle = 
	O_p \left( 
		%\frac{\left( \sum_{k=1}^K \sum_{\ell \in \AR^{(k)}} (\tbeta^{(k)}_\ell \sigma_{\ell j})^2 \right)^{1/2}}{K} 
		\frac{\sigma}{K}  \left[ \sum_{k=1}^K \| \tbeta^{(k)} \|_2^2 \right]^{1/2}
	\right). 
\end{align} 

% term 3
{\em Bounding term 3.} 
We begin by stating a helpful lemma below, the proof of which can be found in Appendix~\ref{sec:lemma_proof_2}. 

\begin{lemma} [Lemma F.2 of \cite{agarwal2021synthetic}] \label{lemma:param_est.2}
Let the setup of Theorem~\ref{thm:consistency} hold. Then for every $k$, the following holds with probability at least $1 - O((|\AC^{(k)}| |\AR^{(k)}|)^{-10})$: 
\begin{align}  
	\| \Delta^{(k)} \|_2^2 ~\le %O \left( 
			C(\sigma) \frac{r^{(k)}\| \tbeta^{(k)} \|_2^2 \log(|\AC^{(k)}| |\AR^{(k)}|)} {\min\{ |\AC^{(k)}|, |\AR^{(k)}| \}},
\end{align} 
where $C(\sigma)$ is a constant that only depends only $\sigma$.
\end{lemma} 
By Lemma~\ref{lemma:param_est.2}, it immediately follows that 
%By Corollary 5.1 of \cite{agarwal2020principal}, $\Ec_k$ occurs w.p. at least $1 - O((|\AC^{(k)}| |\AR^{(k)}|)^{-10})$, which implies that
\begin{align} \label{eq:4} 
	\Delta^{(k)} = O_p \left( \frac{ (r^{(k)})^{1/2}\| \tbeta^{(k)} \|_2 \log^{1/2}(|\AC^{(k)}| |\AR^{(k)}|)} {\min\{ |\AC^{(k)}|^{1/2}, |\AR^{(k)}|^{1/2} \}} \right). 
\end{align} 
For every $k$, we define the event $\Ec_k$ as 
\begin{align}
	\Ec_k = \left\{
		\| \Delta^{(k)} \|_2^2 ~\le %O \left( 
			\frac{ r^{(k)} \| \tbeta^{(k)} \|_2^2 \log(|\AC^{(k)}| |\AR^{(k)}|)} {\min\{ |\AC^{(k)}|, |\AR^{(k)}| \}} 
		%\right)
	\right\}. 
\end{align} 
For ease of notation, let $\Ec_\sharp = \cap_{k=1}^K \Ec_k$. 
Next, we define the event 
\begin{align}
	\Ec_\flat = \left\{
		\frac{1}{K} \sum_{k=1}^K \langle \varepsilon^{(k)}, \Delta^{(k)} \rangle 
		= O\left(
			\frac{\sigma}{K} 
			\left[
				\sum_{k=1}^K \frac{r^{(k)} \| \tbeta^{(k)} \|_2^2 \log(|\AC^{(k)}| |\AR^{(k)}|)}{\min\{|\AC^{(k)}|, |\AR^{(k)}|\}}
			\right]^{1/2}
			%\frac{C(\|\tbeta^{(k)}\|_2) \log^{1/2}(|\AC^{(k)}| |\AR^{(k)}|)}{\sqrt{K} \cdot \min\{|\AC^{(k)}|^{1/2}, |\AR^{(k)}|^{1/2} \}}
		\right)
	\right\}. 
\end{align} 
Now, condition on $\Ec_\sharp$. By Assumption~\ref{assump:subg}, we have for every $k$, 
\begin{align}
	\Ex[\langle \varepsilon^{(k)}, \Delta^{(k)}\rangle] &= 0
	\\ \text{Var}(\langle \varepsilon^{(k)}, \Delta^{(k)}\rangle) &= \sum_{\ell \in \AR^{(k)}}  \sigma_{\ell j}^2 ~ (\hbeta^{(k)}_\ell - \tbeta^{(k)}_\ell)^2 \le \sigma^2 \| \Delta^{(k)} \|_2^2. \label{eq:delta_k_product_bound} 
\end{align}
Given that $\langle \varepsilon^{(k)}, \Delta^{(k)}\rangle$ are independent across $k$, Lemmas~\ref{lemma:a_hoeffding},~\ref{lemma:param_est.2}, and \eqref{eq:delta_k_product_bound} imply $\Ec_\flat | \Ec_\sharp$ occurs w.h.p. 
Further, 
\begin{align} \label{eq:5} 
	\Pb(\Ec_\flat) = \Pb(\Ec_\flat | \Ec_\sharp) \Pb(\Ec_\sharp) + \Pb(\Ec_\flat | \Ec_\sharp^c) \Pb(\Ec_\sharp^c) \ge \Pb(\Ec_\flat | \Ec_\sharp) \Pb(\Ec_\sharp). 
\end{align}
Applying the union bound and DeMorgan's Law, we obtain
\begin{align}
	\Pb(\Ec_\sharp^c) = \Pb(\cup_{k=1}^K \Ec_k^c) \le \sum_{k=1}^K \Pb(\Ec_k^c) \le K \max_k \Pb(\Ec_k^c) 
	= O\left(\frac{K}{\min_k  |\AC^{(k)}|^{10} |\AR^{(k)}|^{10}} \right),
\end{align}
where the final equality follows from Lemma~\ref{lemma:param_est.2}. 
From our condition on $K = o(\min_k |\AC^{(k)}|^{10} |\AR^{(k)}|^{10})$, we have that $\Ec_\sharp$ occurs w.h.p.
Since both $\Ec_\sharp$ and $\Ec_\flat | \Ec_\sharp$ occur w.h.p., it follows from \eqref{eq:5} that $\Ec_\flat$ then occurs w.h.p. 
Therefore, 
\begin{align} 
 \frac{1}{K} \sum_{k=1}^K \langle \varepsilon^{(k)}, \Delta^{(k)} \rangle 
		&= O_p\left(
			\frac{\sigma}{K} 
			\left[
				\sum_{k=1}^K \frac{ r^{(k)}\| \tbeta^{(k)} \|_2^2 \log(|\AC^{(k)}| |\AR^{(k)}|)}{\min\{|\AC^{(k)}|, |\AR^{(k)}|\}}
			\right]^{1/2}
		\right),
		\\&= O_p\left(
			\frac{\sigma}{K} 
				\sum_{k=1}^K \frac{ (r^{(k)})^{1/2} \| \tbeta^{(k)} \|_2 \log^{1/2}(|\AC^{(k)}| |\AR^{(k)}|)}{\min\{|\AC^{(k)}|^{1/2}, |\AR^{(k)}|^{1/2}\}} \label{eq:lemma.3}
		\right).
\end{align}
%
% where the second inequality follows from the fact that $\sqrt{A + B} \le \sqrt{A} + \sqrt{B}$.

% collecting terms
{\em Collecting terms.} 
Incorporating \eqref{eq:lemma.1}, \eqref{eq:lemma.2}, \eqref{eq:lemma.3} into \eqref{eq:1}, and simplifying yields
\begin{align}
	\hA_{ij} - A_{ij} 
	&= O_p
	\left( 
	\frac{1}{K}\left\{
	\sum_{k=1}^K \frac{(r^{(k)})^{1/2}}{ |\AC^{(k)}|^{1/4}} 
	+  \sum_{k=1}^K \frac{(r^{(k)})^{3/2} \| \tbeta^{(k)}\|_1 \log^{1/2}(|\AC^{(k)}| |\AR^{(k)}|)} { \min\{ |\AC^{(k)}|^{1/2}, |\AR^{(k)}|^{1/2} \}}
	+ \left[ \sum_{k=1}^K \| \tbeta^{(k)} \|_2^2 \right]^{1/2} 
\right\} \right).
\end{align} 
This concludes the proof.

% LEMMA 1 PROOF
\subsection{Proof of Lemma~\ref{lemma:param_est}} \label{sec:lemma_proof_1}
The result is immediate from Lemma G.1 of \cite{agarwal2021synthetic} after adapting the notation used in \cite{agarwal2021synthetic} to that used in this paper. 
For every $k$, let $Y_{\text{pre},n} = q$, $\Ex[\bY_{\text{pre}, \Ic^{(d)}}] = \Ex[\bS^{(k)}]$, $\bY_{\text{pre}, \Ic^{(d)}} = \bS^{(k)}$, $\bV_{\text{pre}} = \bU^{(k)}$, $\hw^{(n,d)} = \hbeta^{(k)}$, $\tw^{(n,d)} = \tbeta^{(k)}$, where $(Y_{\text{pre},n},  \Ex[\bY_{\text{pre}, \Ic^{(d)}}], \bY_{\text{pre}, \Ic^{(d)}}, \bV_{\text{pre}},  \hw^{(n,d)}, \tw^{(n,d)})$ are the notations used in \cite{agarwal2021synthetic}.

% LEMMA 2 PROOF
\subsection{Proof of Lemma~\ref{lemma:param_est.2}} \label{sec:lemma_proof_2} 
The result is immediate from Lemma F.2 of \cite{agarwal2021synthetic} after adapting the notation used in \cite{agarwal2021synthetic} to that used in this paper. 
For every $k$, let $Y_{\text{pre},n} = q$, $\Ex[\bY_{\text{pre}, \Ic^{(d)}}] = \Ex[\bS^{(k)}]$, $\bY_{\text{pre}, \Ic^{(d)}} = \bS^{(k)}$, $\bV_{\text{pre}} = \bU^{(k)}$, $\hw^{(n,d)} = \hbeta^{(k)}$, $\tw^{(n,d)} = \tbeta^{(k)}$, where $(Y_{\text{pre},n},  \Ex[\bY_{\text{pre}, \Ic^{(d)}}], \bY_{\text{pre}, \Ic^{(d)}}, \bV_{\text{pre}},  \hw^{(n,d)}, \tw^{(n,d)})$ are the notations used in \cite{agarwal2021synthetic}. 
%
% In addition, we use the fact that $\Ex[\bY_{\text{pre}, \Ic^{(d)}}] \tw^{(n,d)}$ is equivalent to $\Ex[q] = \Ex[\bS^{(k)}] \tbeta^{(k)}$, which follows from Assumptions~\ref{assump:LFM}, \ref{assump:linear_span}, \ref{assump:mean_ind}, and \ref{assump:subspace}. 

\section{Proof of Proposition \ref{prop:MCAR}}
%
% For the proof $c, C > 0$ denote absolute constants, whose value can change between lines, and even within a line.
%
First, let us consider recovery of entry $(i,j) = (1,1)$.
We let $C > 0$ denote an absolute constant.
Define parameter $Q \ge 1$.
Excluding row $1$ and column $1$, partition the remaining $(L - 1)$ rows and $(L - 1)$ columns into $(L - 1) / Q$ mutually exclusive blocks each of size $Q + 1$.
In particular, $\bM^{(1, 1)}_\ell \in \Rb^{Q + 1 \times Q + 1}$ for $\ell \in [(L - 1) / Q]$ corresponds to the sub-matrix induced by selecting only rows $\{1, (\ell - 1) Q + 2, \dots, \ell Q + 1 \}$ and columns $\{1, (\ell - 1) Q + 2, \dots, \ell Q + 1\}$.
Let $\mathds{1}^{(1, 1)}_{\ell}$ be a binary r.v. which is equal to $1$ if all entries in the sub-matrix $\bM^{(1, 1)}_\ell$ not including $(1, 1)$ are revealed (i.e., we do not condition on whether $(1, 1)$ is revealed or missing).

Define the event 
$
\Ec_{(1, 1)} := \{\mathds{1}^{(1, 1)}_{\ell} = 0: \forall \ \ell \in [(L - 1) / Q]\},
$
i.e., $\Ec_{(1, 1)}$ is the event that none of the $(L - 1) / Q$ sub-matrices $\bM^{(1, 1)}_\ell$ are fully revealed.
Note $\mathds{1}^{(1, 1)}_{\ell}$ is equal to $1$ with probability $p^{Q^2 + 2Q} \ge p^{2Q^2} =: q$.
Observe that $\mathds{1}^{(1, 1)}_{\ell}$ and $\mathds{1}^{(1, 1)}_{\ell'}$ for $\ell \neq \ell^'$ are independent r.v.s.
Then the probability $\Ec_{(1, 1)}$ occurs is at most
$(1 - q)^{(L - 1) / Q} \le \exp^{- \frac{q (L - 1)}{Q}}.$
Note, 
\begin{align}
    \exp^{- \frac{q (L - 1)}{Q}} \le (L - 1)^{-10} \iff q \ge \frac{10 \log((L - 1))  Q}{(L - 1)} \iff p \ge C \left(\frac{\log(L) Q}{L}\right)^{\frac{1}{Q^2}}
\end{align}
To get an additive error of at most $O_p(\delta)$ for $A_{1, 1}$, we require $Q = C^*\delta^{-6}$ by Corollary \ref{cor:consistency}---this can be seen by noting that $Q$ needs to equal the total number of anchor rows which is $N \times K$, where $N$ and $K$ are defined in Corollary \ref{cor:consistency}.
In summary, we have that $A_{i, j} - \hA_{i, j} = O_p(\delta)$ if $Q = C^*\delta^{-6}$ and $\Ec^c_{(1, 1)}$ holds, where $\Ec^c_{(1, 1)}$ occurs with probability at least $1 - (L - 1)^{-10}$ if $p \ge C \left(\frac{\log(L) Q}{L}\right)^{\frac{1}{Q^2}}$.

Now we generalize to any $(i, j)$ pair.
Define $\Ec_{(i, j)}$ analogously to $\Ec_{(1, 1)}$
The difference being that we replace the fixed row and column from $(1, 1)$ to $(i, j)$, and partition the remaining $(L - 1)$ rows and $(L - 1)$ columns to create the matrices $\bM^{(i, j)}_\ell$ for $\ell \in [(L - 1) / Q]$.
$\mathds{1}^{(i, j)}_{\ell}$ is then defined with respect to $\bM^{(i, j)}_\ell$, analogous to the way $\mathds{1}^{(1, 1)}_{\ell}$ is defined with respect to $\bM^{(1, 1)}_\ell$.
To ensure that $A_{i, j} - \hA_{i, j} = O(\delta)$ uniformly for all $(i, j)$, we then require the event $\bigcap_{(i, j) \in [L] \times [L]} \Ec^c_{(i, j)}$ to hold with $Q = C^*\delta^{-6}$ as before. 
Appealing to the definition of $p$ in statement of Proposition \ref{prop:MCAR}, this occurs with probability, 
\begin{align}
\Pb( \bigcap_{(i, j) \in [L] \times [L]} \Ec^c_{(i, j)}) 
&= 1 - \Pb( \bigcup_{(i, j) \in [L] \times [L]} \Ec_{(i, j)}) 
\\ &\ge 1 - \sum_{(i, j) \in [L] \times [L]}\Pb(\Ec_{(i, j)})
\\ &\ge 1 - \frac{C}{L^8}.
\end{align}
This completes the proof.
\section{Proof of Theorem~\ref{thm:normality}}
For ease of notation, we suppress the conditioning on $\Ec$ for the remainder of the proof.
To begin, we scale the left-hand side (LHS) of \eqref{eq:1} by
\begin{align}\label{eq:scaling}
\frac{K}{\sqrt{\sum^K_{k = 1} \left(\tilde{\sigma}^{(k)}\right)^2}} 
\end{align}
and analyze each of the three resulting terms on the right-hand side (RHS) separately.

% term 1
{\em Bounding term 1.} 
To address the first term, we scale \eqref{eq:lemma.1} by \eqref{eq:scaling} and recall our assumption on $K$ given by \eqref{eq:normality_cond1}. 
We then obtain
\begin{align} \label{eq:normality.1} 
	\frac{1}{\sqrt{\sum^K_{k = 1} \left(\tilde{\sigma}^{(k)}\right)^2}} \sum_{k=1}^K \langle \Ex[x^{(k)}], \Pc_{U^{(k)}} \Delta^{(k)} \rangle = o_p(1). 
\end{align} 

% {\color{red}
% %
% \begin{align}
%     O_p\left( 
% 	% + \max_k
% 	 \frac{1}{\sqrt{K}}\sum_{k=1}^K \frac{1}{ |\AC^{(k)}|^{1/4}}  
% 	 + \frac{\| \tbeta^{(k)}\|_1 \log^{1/2}(|\AC^{(k)}| |\AR^{(k)}|)} { \min\{ |\AC^{(k)}|^{1/2}, |\AR^{(k)}|^{1/2} \}}\right)
% \end{align}
% Need
% \begin{align}
%     \sum_{k=1}^K \left( \frac{1}{ |\AC^{(k)}|^{1/4}} + \frac{\| \tbeta^{(k)}\|_1 \log^{1/2}(|\AC^{(k)}| |\AR^{(k)}|)} { \min\{ |\AC^{(k)}|^{1/2}, |\AR^{(k)}|^{1/2} \}}\right)  = o(\sqrt{K})
% \end{align}
% }

% term 2
{\em Bounding term 2.}
Since $\langle \varepsilon^{(k)}, \tbeta^{(k)} \rangle$ are independent across $k$, the Lindeberg-L\'evy Central Limit Theorem and \eqref{eq:variance_per_k} yields
\begin{align} \label{eq:normality.2} 
	\frac{\sum_{k=1}^K \langle \varepsilon^{(k)}, \tbeta^{(k)} \rangle}
	{\sqrt{\sum^K_{k = 1} \left(\tilde{\sigma}^{(k)}\right)^2}}
	\xrightarrow{d} \mathcal{N}(0, 1). 
\end{align} 

% term 3
{\em Bounding term 3.}
Next, we scale \eqref{eq:lemma.3} by \eqref{eq:scaling} and recall our assumption on $K$.
This yields
\begin{align} \label{eq:normality.3} 
	\frac{1}{\sqrt{\sum^K_{k = 1} \left(\tilde{\sigma}^{(k)}\right)^2}} \sum_{k=1}^K \langle \varepsilon^{(k)}, \Delta^{(k)} \rangle = o_p(1). 
\end{align} 

% collecting terms
{\em Collecting terms.} 
From \eqref{eq:normality.1}, \eqref{eq:normality.2}, and \eqref{eq:normality.3}, we conclude 
\begin{align}
	\frac{K(\hA_{ij} - A_{ij})} {\sqrt{\sum^K_{k = 1} \left(\tilde{\sigma}^{(k)}\right)^2}} 
	\xrightarrow{d} \mathcal{N}(0,1).  
\end{align}

\end{document}